\documentclass[aps,prb,reprint,showpacs,superscriptaddress, longbibliography]{revtex4-1}
\pdfoutput=1
\usepackage{mathtools}
\usepackage{graphicx}
\usepackage{amsmath}
\usepackage{amsfonts}
\usepackage{color}
\usepackage{hyperref}

\newcommand{\ud}{\mathrm{d}}
\newcommand{\Ket}[1]{| #1 \rangle}

\newcommand{\Braket}[1]{\langle #1 \rangle}
\newcommand{\Comm}[2]{\left[#1, #2\right]}
\newcommand{\V}[2]{\phi_{#1}(#2)}
\newcommand{\Vs}[1]{\phi_{#1}}
\newcommand{\Vi}[1]{\phi_{s_{#1}}(z_{#1})}
\newcommand{\Ja}[1]{J^{a_{#1}}_{-1}}
\newcommand{\Jpsi}{\psi_{a_k \dots a_1}}
\newcommand{\Gpsi}{\psi_0}

\newcommand{\Hf}{\mathcal{H}}
\newcommand{\It}[2]{u^{#1}_{#2}}
\newcommand{\Ia}[1]{u^{a_{#1}}_{-1}}
\newcommand{\Co}[2]{\mathcal{C}^{#1}_{#2}}

\newcommand{\PhiString}{\Phi_{\mathbf{s}}(\mathbf{z})}

\newcommand{\cellblock}[1]{%
  \begin{tabular}[c]{@{}l@{}}#1\end{tabular}}
\newcommand{\JString}{J^{a_k}_{-1}\dots J^{a_1}_{-1}}
\newcommand{\MPQ}{Max-Planck-Institut f\"ur Quantenoptik, Hans-Kopfermann-Str.\ 1, D-85748 Garching, Germany}

\makeatletter
\newcommand{\redboxed}[1]{\textcolor{red}{%
  \fbox{\normalcolor\m@th$\displaystyle#1$}}}
\makeatother
\def\vep{\varepsilon}

\begin{document}

\title{Excited States in Spin Chains from Conformal Blocks}

\begin{abstract}
We develop a method of constructing excited states in one dimensional
spin chains which are derived from the $SU(2)_1$ Wess-Zumino-Witten
Conformal Field Theory (CFT) using a parent Hamiltonian approach.
The resulting systems are equivalent to the Haldane-Shastry model.
In our ansatz, correlation functions between primary fields correspond
to the ground state of the spin system, whereas excited states are
obtained by insertion of descendant fields. Our construction is based
on the current algebra of the CFT and emphasizes the close relation
between the spectrum of the spin system and the underlying CFT.
This general structure might imply that the method could be applied
to a wider range of model systems.
\end{abstract}

\pacs{75.10.Pq, 71.10.Li, 11.25.Hf}
\keywords{excited states; spin chains; Haldane-Shastry model; Conformal Field Theory}

\author{Benedikt Herwerth}
\affiliation{\MPQ}
\author{Germ\'an Sierra}
\affiliation{Instituto de F\'isica Te\'orica, UAM-CSIC, Madrid, Spain}
\author{Hong-Hao Tu}
\author{Anne E. B. Nielsen}
\affiliation{\MPQ}

\maketitle
% \tableofcontents

\section{Introduction}
\label{SEC:Introduction}
The description of quantum many body systems is an intrinsically
hard problem due to the dimensionality of the Hilbert space,
which depends exponentially on the system size.
Despite the impressive success of numerical and approximate
techniques like Exact Diagonalization \cite{Laeuchli2011},
Quantum Monte Carlo \cite{Metropolis1953, Handscomb1962},
and the Density Matrix Renormalization Group \cite{White1992}, exactly solvable
models remain essential in studying the physics of quantum many body
systems. An analytical solution is not only indispensable for benchmarking
approximate methods, it can also elucidate the general structure of solutions
and underlying physical principles.
With the experimental advances in cooling and
controlling the interactions of atoms, interesting model
systems that have a closed theoretical solution
may even be engineered and studied in the laboratory.

Given the complexity of a generic quantum many body problem, the
direct construction of exactly solvable models appears as an appealing
approach. In the past years, $1+1$ dimensional Conformal Field Theory
(CFT) has proven to be a powerful tool to construct model Hamiltonians
and corresponding ground state wave functions for
continuum and lattice quantum Hall models
\cite{Moore1991, Cirac2010, Nielsen2013a}.  In this approach, a CFT is taken as
the starting point to derive a quantum many body system. It can thus
contribute to a systematic understanding of systems that admit a
description in terms of a scale invariant theory, such as critical
systems or systems with topological order that have a gapless edge
spectrum.  Theories for which such an analysis was carried out include
the $SU(2)_k$ \cite{Nielsen2011}, $SO(n)_1$ \cite{Tu2013c}, $U(1)_q$
\cite{Tu2014b}, and $SU(n)_1$ \cite{Tu2014, Bondesan2014}
Wess-Zumino-Witten (WZW) theories.  Another correspondence between CFT
and a class of continuum and lattice quantum systems in two dimensions was
studied in Ref. \onlinecite{Ardonne2004}.  In these models, the ground
state exhibits a $z=2$ Lifshitz scale invariance.

In this paper, we describe a method of constructing excited states
of a spin system that is derived from a CFT.
Correlation functions of fields in the CFT are interpreted as wave functions of states in the
spin system.
Following earlier studies \cite{Nielsen2011, Nielsen2012},
we consider the $SU(2)_1$ WZW model and
define a parent Hamiltonian for the state that
corresponds to the correlator of $N$ primary fields.
In the case of periodic boundary conditions in one
dimension, which we focus on here, the resulting spin system is equivalent to
the Haldane-Shastry model \cite{Haldane1988, Shastry1987}.

Based on solutions to the Calogero-Sutherland model\cite{Calogero1969,
Calogero1971, Sutherland1971, Sutherland1971a}, excited states of the
Haldane-Shastry Hamiltonian were first constructed as polynomials in
the particle basis\cite{Haldane1988, Haldane1991}.  It was realized
later\cite{Haldane1992, Talstra1995PhD} that these states are the
highest weight states of the Yangian algebra, a hidden symmetry of the
Haldane-Shastry Hamiltonian. Given the close relation between the
Haldane-Shastry model and the $SU(2)_1$ WZW CFT, it was conjectured in
Ref.~\onlinecite{Haldane1992} that there should be a correspondence
between the states in the CFT and those of the spin chain. Here we
show that this is indeed the case. Since our ansatz is
based on the $SU(2)$ currents, it is manifestly $SU(2)$ invariant.
We relate our results both to the Yangian highest weight states and to
the spinon basis of CFT states \cite{Bouwknegt1994}.

Our ansatz mirrors the structure of the CFT, with the ground state
corresponding to the CFT vacuum and the excited states corresponding
to CFT descendant states.  More precisely, we construct the excited
states from the ground state by insertion of current operator modes
into correlation functions of primary fields.  We show that the
Hamiltonian of the spin system is block diagonal in these states, with
each block corresponding to a fixed number of inserted current
operators.  This allows us to obtain excited states by successively
adding current operators and diagonalizing the blocks. As the
Hamiltonian, the Yangian is closed in the subspaces of states with a certain number
of current operators. This implies that one can construct representations
of the Yangian algebra in these subspaces. We explicitly construct
the highest weight states for the analytically obtained eigenstates. Our ansatz for
the excited states and its relation to descendant states of the CFT
is summarized in Table \ref{table:structure-of-excited-states}.

\begin{table}[htb]
  \caption{\label{table:structure-of-excited-states}
    Structure of the states in the CFT and the spin system.  The
    ground state of the spin system is constructed from the product of $N$
    primary fields $\PhiString = \phi_{s_1}(z_1) \dots \phi_{s_N}(z_N)$
    and corresponds to the CFT vacuum $\Ket{0}$. Excited states are
    constructed from the ground state by insertion of current operator
    modes $J^{a_k}_{-1}\dots J^{a_1}_{-1}$.
  }
\centering
\begin{tabular}{llcl}
 & CFT & &  Spin system\\
\hline
\cellblock{Ground\\state} & $\Ket{0}$ & $\leftrightarrow$
         & $\textcolor{blue}{\langle 0| \PhiString}  | 0 \rangle$\\
 & $\downarrow$ &  & $\downarrow$\\
\cellblock{Excited\\states} & $\textcolor{red}{(J^{a_k}_{-1} \dots J^{a_1}_{-1})(0)} \Ket{0}$
       & $\leftrightarrow$ & $\textcolor{blue}{\langle 0| \PhiString} \textcolor{red}{(J^{a_k}_{-1} \dots J^{a_1}_{-1})(0)} | 0 \rangle$
\end{tabular}
\end{table}

We perform the diagonalization of the Hamiltonian analytically for up
to eight current operators, construct Yangian highest weight states
in the obtained eigenstates, and confirm numerically for small system
sizes that the complete spectrum can be obtained in this way.  We show
that this construction can be done both for an even and for an odd
number of spins in the chain.

This paper is structured as follows: In Section
\ref{SEC:states-from-conformal-blocks}, we review some properties of
the $SU(2)_1$ WZW model and describe the states that form the basis of
our construction. We introduce the parent Hamiltonian in Section
\ref{SEC:Parent-Hamiltonian-and-Spectrum} and construct excited
states, both analytically (Section \ref{SEC:Eigenstates-Analytically})
and numerically (Section \ref{SEC:Numerics}). We conclude in Section
\ref{SEC:Conclusion}.

\section{States From Conformal Blocks}
\label{SEC:states-from-conformal-blocks}
In this section, we briefly review some properties of the $SU(2)_1$ WZW model
and describe the correspondence between conformal blocks and spin system
wave functions.

In addition to the identity, the model has one primary field
with scaling dimension $h = 1/4$, the vertex operator $\V{s}{z}$. It
can be constructed from the chiral part $\varphi(z)$ of a free, massless boson as
\begin{align}
\label{eq:vertex-operator-from-free-boson}
\V{s}{z} & = e^{i \pi (q-1) (s+1) / 2} :e^{i s \varphi(z)/ \sqrt{2}}:.
\end{align}
Here $s=\pm1$ corresponds to the two components of the vertex
operator, and the colons denote normal ordering.  (The holomorphic
field $\phi_s(z)$ has an anti-holomorphic counterpart
$\bar{\phi}_s(\bar{z})$.  In the following we only consider the
holomorphic sector.)  The value $q \in \{0, 1\}$ corresponds to the
two sectors of the CFT: $q=0$ if the operator $\V{s}{z}$ acts on a
state that has an even number of $h=1/4$ primary fields and $q=1$ for
a state with an odd number, respectively.

In addition to conformal invariance, the $SU(2)_1$ WZW model has an $SU(2)$
symmetry, which is generated by the current operator $J^a(z)$.
Its Laurent expansion defines modes $J^a_n$,
\begin{align}
J^a(z) &= \sum^{\infty}_{n = -\infty} z^{-n - 1} J^a_n,
\end{align}
where $a \in \{x, y, z\}$.
They satisfy the Kac-Moody algebra \cite{Knizhnik1984}
\begin{align}
\label{eq:Kac-Moody-current-algebra}
\Comm{J^a_m}{J^b_n} &= i \vep_{abc} J^c_{m+n} + \frac{m}{2} \delta_{ab} \delta_{m+n, 0},
\end{align}
with $\vep_{abc}$ being the Levi-Civita symbol and $\delta_{ab}$ the Kronecker delta.

A primary field $\V{s}{z}$ transforms covariantly with respect to conformal
and $SU(2)$ transformations.  The latter is expressed by
the operator product expansion (OPE) between the current $J^a(z)$
and $\V{s}{z}$ \cite{DiFrancesco1997},
\begin{align}
\label{eq:ope-j-with-vertex-op}
J^a(z) \V{s}{w} &\sim -\sum_{s'}\frac{(t^a)_{s s'}}{z-w} \V{s'}{w}.
\end{align}
Here $t^a$ are the $SU(2)$ spin operators. They are related
to the Pauli matrices $\sigma^a$ by $t^a = \sigma^a / 2$.

The modes of the current operator
give rise to a tower of descendant states
\begin{align}
\label{eq:descendants-in-CFT}
&(J^{a_k}_{-1} \dots J^{a_1}_{-1})(0) \Ket{0},
\end{align}
with $\Ket{0}$ being the CFT vacuum. States of this form build up the spectrum of the CFT \cite{Gepner1986}.
Note that it suffices to consider states obtained from the $n=-1$ mode of
$J^a(z)$. This is so because one can successively rewrite a higher
order mode $J^a_{-n}$, $n > 0$, in terms of the lower order modes $J^a_{-n+1}$
and $J^a_{-1}$ by means of the Kac-Moody algebra (cf. Eq. \eqref{eq:Kac-Moody-current-algebra}),
\begin{align}
\label{eq:elim-higher-order-J}
J^a_{-n} &= \frac{i}{2} \vep_{abc} \Comm{J^{c}_{-1}}{J^{b}_{-n+1}}, n \neq 0.
\end{align}

In this work, we show that the spectrum of the spin systems
organizes in the same way in terms of current operators.
The excited states we obtain are linear combinations
of conformal blocks containing current operator
modes $J^{a_k}_{-1} \dots J^{a_1}_{-1}$.
We give a summary of these states in Table \ref{table:summary-of-states}
and describe the construction of states from conformal blocks in the next subsections.
\begin{table}[htb]
  \caption{\label{table:summary-of-states}
    Summary of the different towers of states obtained by insertion of
current operator modes.  $\PhiString$ denotes the product of $N$
primary fields, $\PhiString = \phi_{s_1}(z_1) \dots \phi_{s_N}(z_N)$.
Using the OPE between $\phi_s(z)$ and $J^a(z)$, the wave functions for
these states can be written as the application of $k$ Fourier transformed
spin operators to the state without current operators, as
explained in Section \ref{SEC:States-from-Vertex-and-Current-Operators}.
  }
\centering
\begin{tabular}{lll}
  & Tower of states & See Eq. \\
\hline
$N$ even & $\Braket{ \PhiString (\JString)(0)}$ & \eqref{eq:def-J-states} \\
$N$ even & $\Braket{ \phi_{s_\infty}(\infty) \PhiString (\JString \phi_{s_0})(0)}$ & \eqref{eq:def-tower-above-psi0inf-sgl} \\
$N$ odd & $\Braket{\phi_{s_\infty}(\infty) \PhiString (\JString)(0)}$ & \eqref{eq:tower-above-psi0odd} \\
$N$ odd & $\Braket{\PhiString (\JString \phi_{s_0})(0)}$ & \eqref{eq:alpha0odd-tower}
\end{tabular}
\end{table}

\subsection{State Obtained from a String of Vertex Operators}
\label{SEC:State-from-Vertex-Operators}

We consider an even number of vertex operators $\Vi{i}$ ($i = 1, \dots, N$)
that each transform under a representation of $SU(2)$ generated by
spin operators $t^a_i$, as expressed by the OPE of
Eq. \eqref{eq:ope-j-with-vertex-op}. The key idea is to view
the correlation function of vertex operators in the CFT
as the wave function of a system of spin $1/2$
degrees of freedom on a lattice,
\begin{align}
\label{eq:def-ground-state}
\Ket{\Gpsi} &= \sum_{s_1 \dots s_N} \psi_0(s_1, \dots, s_N) \Ket{s_1, \dots, s_N},
\intertext{where $\Gpsi(s_1, \dots, s_N)$ is given by}
\Gpsi(s_1, \dots, s_N) &= \Braket{\V{s_1}{z_1} \dots \V{s_N}{z_N}}.
\end{align}
Here,
\begin{align}
\label{eq:vertex-operators-with-sign-factor}
\Vi{j} &= e^{\pi i (j-1) (s_j + 1) / 2} :e^{i s_j \varphi(z_j)/ \sqrt{2}}:,
\end{align}
and $\Ket{s_1, \dots, s_N}$ with $s_i = \pm 1$ is the tensor product
of eigenstates $\Ket{s_i }$ of the $z$-component of the spin operator
$t^z_i$.  The coordinates $z_i$ define the lattice positions in the
complex plane and are kept fixed.  In contrast to the continuum case,
there is no spatial degree of freedom in the basis states $\Ket{s_1,
\dots, s_N}$ and therefore no integral over the positions in
Eq. \eqref{eq:def-ground-state}. This is why we use the notation
$\Gpsi(s_1, \dots s_N)$ without the coordinates $z_i$ for the spin
wave function.

The correlation function of $N$ vertex operators is given by
\cite{DiFrancesco1997}
\begin{align}
\label{eq:vertex-op-correlator}
  \Gpsi(s_1, \dots, s_N) &= \Braket{\V{s_1}{z_1} \dots\V{s_N}{z_N}} \\
  &= \delta_{\mathbf{s}} \chi_{\mathbf{s}} \prod_{i < j}^{N} (z_i - z_j)^{s_i s_j / 2}\notag,
\end{align}
where $\delta_{\mathbf{s}}$ is $1$ if $\sum_{i=1}^N s_i = 0$ and $0$
otherwise. $\chi_\mathbf{s}$ is the Marshall sign factor,
\begin{align}
  \label{eq:marshall-sign-factor}
  \chi_{\mathbf{s}} &= \prod_{p=1}^{N} e^{i \pi (p-1) (s_p+1)/2},
\end{align}
which ensures that the state $\Gpsi$ is a spin singlet\cite{Cirac2010, Nielsen2012},
\begin{align}
  \label{eq:psi0-singlet-property}
  T^a \psi_0 &= 0, \quad T^a = \sum_{i=1}^{N} t^a_i.
\end{align}
Note that the condition $\delta_{\mathbf{s}}$ requires the
number of primary fields $\V{s_j}{z_j}$ in the correlator to be even.

The positions $z_j$ can, in principle, assume any value on the complex
plane. In the following we mostly consider $N$ spins uniformly distributed
on the circle,
\begin{align}
\label{eq:zi-on-unit-circle}
z_j &= z^j, \quad\text{with } z = e^{2 \pi i/N}.
\end{align}
In this case, the state $\psi_0$ has a momentum of $\pi$ if $N/2$ is odd and $0$
if $N/2$ is even\cite{Bernevig2001}, see also Appendix \ref{SEC:Lattice-momentum}.

\subsection{States Obtained from Vertex and Current Operators}
\label{SEC:States-from-Vertex-and-Current-Operators}
In analogy to the tower of CFT descendant states $(J^{a_k}_{-1} \dots J^{a_1}_{-1})(0) \Ket{0}$,
we define a tower of spin states by insertion of
modes of the current operator $J^a(z)$ into a correlation function
of vertex operators,
\begin{align}
  \label{eq:def-J-states}
  &\Ket{\Jpsi} = \sum_{s_1 \dots s_N} \Jpsi(s_1, \dots, s_N) \Ket{s_1, \dots, s_N}, \notag \\
  \intertext{where $\Jpsi(s_1, \dots, s_N)$ is given by}
  &\Jpsi(s_1, \dots, s_N) \notag \\
  & \quad= \Braket{\V{s_1}{z_1} \dots \V{s_N}{z_N} (J^{a_k}_{-1} \dots J^{a_1}_{-1})(0)}.
\end{align}

It is possible to obtain the states $\Jpsi$ by applying spin operators to $\Gpsi$.
To see this, we first note that
\begin{equation}
\langle \PhiString (J^a_{-1}B)(0)\rangle =\frac{1}{2\pi i}\oint_0\frac{\ud w}{w}\langle \PhiString J^a(w)B(0)\rangle,
\end{equation}
with $\PhiString \equiv \Vi{1} \dots \Vi{N}$ and $B$ being an arbitrary operator.
Applying this relation to the definition of $\Jpsi$ we obtain
\begin{align}
\label{eq:derivation_of_Jpsi}
&\Jpsi(s_1, \dots, s_N) \notag \\
&= \Braket{\Vi{1} \dots \Vi{N} (J^{a_k}_{-1} \dots J^{a_1}_{-1})(0)} \notag \\
&= \frac{1}{2\pi i}\oint_0\frac{\ud w}{w}\langle \PhiString J^{a_k}(w)(J^{a_{k-1}}_{-1}\ldots J^{a_1}_{-1})(0)\rangle \notag \\
&=-\frac{1}{2\pi i}\sum_{j=1}^N\oint_{z_j}\frac{\ud w}{w}\langle \PhiString J^{a_k}(w)(J^{a_{k-1}}_{-1}\ldots J^{a_1}_{-1})(0)\rangle. \\
\intertext{Using the OPE between a current operator and a primary field (cf. Eq. \eqref{eq:ope-j-with-vertex-op}), we get}
&\Jpsi(s_1, \dots, s_N) \notag \\
&= \frac{1}{2\pi i}\sum_{j=1}^N\oint_{z_j}\frac{\ud w}{w}\frac{t_j^{a_k}}{w-z_j} \langle \PhiString (J^{a_{k-1}}_{-1}\ldots J^{a_1}_{-1})(0)\rangle \notag \\
&= \sum_{j=1}^N\frac{t_j^{a_k}}{z_j} \psi_{a_{k-1}\ldots a_1}.
\end{align}
Successive application of the same argument results in
\begin{align}
\label{eq:result-Jpsi-from-Gpsi}
\Jpsi &= \left(\sum_{j_k=1}^{N} \frac{t^{a_k}_{j_k}}{z_{j_k}}\right) \dots \left(\sum_{j_1=1}^{N} \frac{t^{a_1}_{j_1}}{z_{j_1}}\right) \Gpsi.
\end{align}
If the positions $z_j$ are uniformly distributed on the circle, we can express this
result in terms of Fourier transformed spin operators $u^{a}_{l}$,
\begin{align}
\label{eq:ft-spin-ops-definition}
u^a_{l} &\equiv \sum_{j=1}^{N} t^a_j e^{2 \pi i j l / N}.
\end{align}
With $z=e^{2\pi i /N}$ in Eq. \eqref{eq:result-Jpsi-from-Gpsi}, we have
\begin{align}
\label{eq:def-I-spin-ops}
\Jpsi &= \It{a_k}{-1} \dots \It{a_1}{-1} \Gpsi.
\end{align}
Therefore, each additional insertion of $J^{a}_{-1}$ changes the momentum
of the state by $2 \pi / N$.

\subsection{States with Additional Spins at Zero and Infinity}
\label{SEC:definition-psi-zero-inf}
\begin{figure}[htb]
\centering
\includegraphics[width=.6\linewidth]{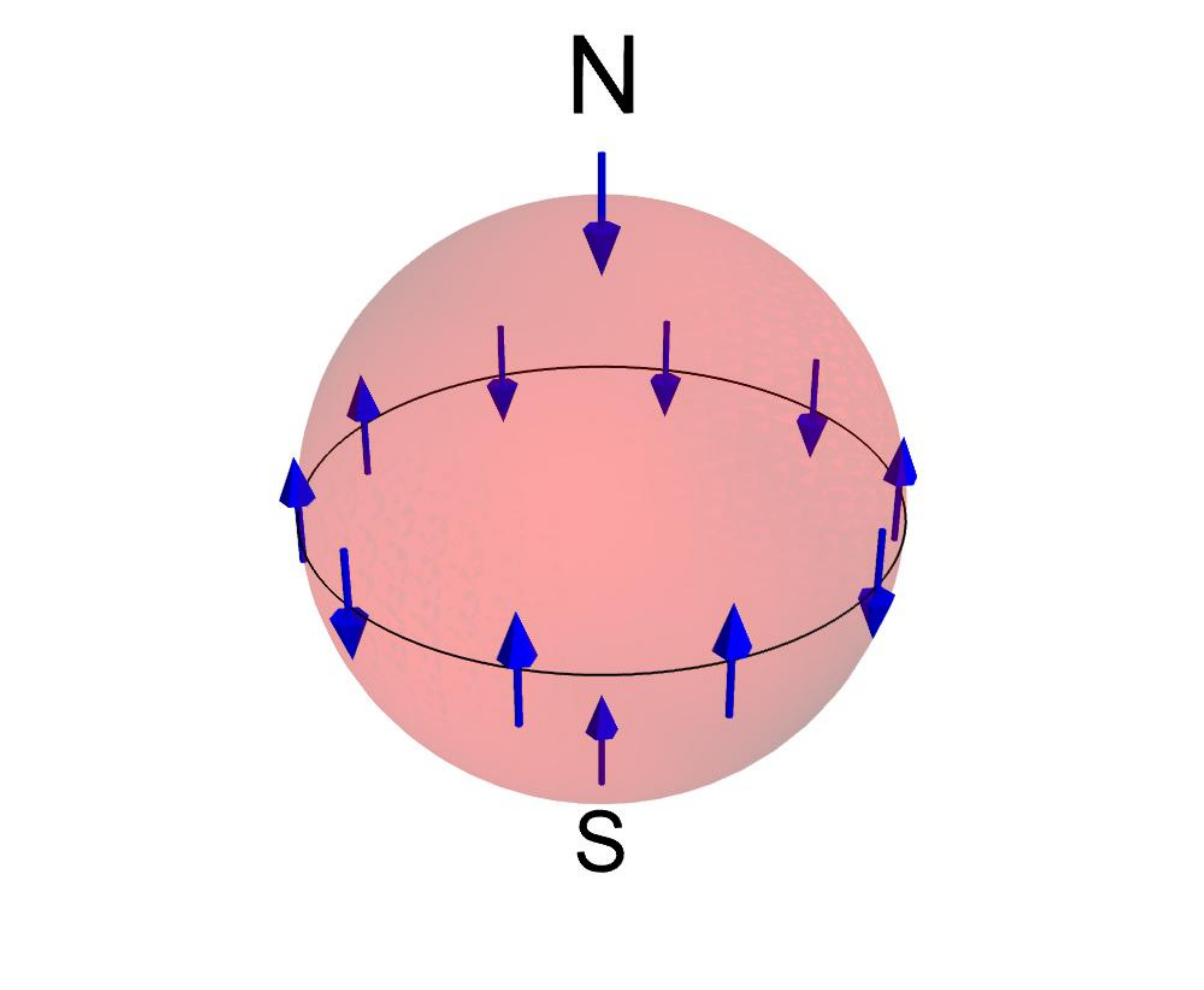}
\caption{\label{fig:Riemann-sphere}
  (Color online) The Riemann sphere with $N$ spins at the equator (unit circle) and two
  additional spins, one at the north pole ($z=0$) and one at the south pole ($z=\infty$).
}
\end{figure}

We define an additional class of states
by inserting two extra vertex operators into the correlator $\Braket{\Vi{1} \dots \Vi{N}}$,
one at $z=0$ and one at $z=\infty$,
\begin{align}
\label{eq:def-0inf-corr}
  &\Ket{\psi^{s_0, s_\infty}_0} = \sum_{s_1, \dots, s_N} \psi_0^{s_0, s_\infty}(s_1, \dots, s_N) \Ket{s_1, \dots, s_N}, \\
  &\psi_0^{s_0, s_\infty}(s_1, \dots, s_N)\notag \\
  &\quad\quad= \Braket{\V{s_\infty}{\infty} \Vi{1} \dots \Vi{N} \V{s_0}{0}}. \notag
\end{align}
In the Riemann sphere picture, the additional spins are added
at the south and north pole, respectively, while the $N$ spins at the
unit circle are located at the equator (see Fig. \ref{fig:Riemann-sphere}).

The wave function is
\begin{align}
  \label{eq:wave-function-alpha0}
  \psi^{s_0, s_\infty}_0(s_1, \dots, s_N) &\propto \delta_{\bar{\mathbf{s}}} (-1)^{s_0 (1-s_\infty)/2} \chi_{\mathbf{s}} \prod_{n=1}^{N} z_n^{s_0 s_n / 2} \notag \\
  &\quad \times \prod_{n < m}^N (z_n - z_m)^{s_n s_m / 2},
\end{align}
where $\delta_{\bar{\mathbf{s}}} = 1$ for $s_0 + s_\infty + \sum_{i=1}^N s_i = 0$ and
$\delta_{\bar{\mathbf{s}}} = 0$ otherwise.

Note that the extra fields inserted at zero and infinity
are primary fields. If we insert additional current operators,
we generate descendant states. We thus define
a tower of states on top of $\psi^{s_0 s_\infty}_0$,
\begin{align}
\label{eq:def-tower-above-psi0inf-sgl}
  &\Jpsi^{s_0, s_\infty}(s_1, \dots, s_N) \\
  &= \Braket{\V{s_\infty}{\infty} \Vi{1} \dots \Vi{N} (J^{a_k}_{-1} \dots J^{a_1}_{-1} \Vs{s_0})(0)}. \notag
\end{align}
This ansatz thus corresponds to
the tower of CFT descendant states $(J^{a_k}_{-1} \dots J^{a_1}_{-1} \phi_s)(0) \Ket{0}$.

An argument similar to the one given for $\Jpsi$ shows that
\begin{align}
\label{eq:Jpsi-zero-inf-from-spin-operators}
  \Jpsi^{s_0, s_\infty} &= \left(\frac{t^{a_k}_{\infty}}{z_\infty} + \sum_{j_k=1}^{N} \frac{t^{a_k}_{j_k}}{z_{j_k}}\right) \dots \notag \\
&\quad \times \left(\frac{t^{a_1}_{\infty}}{z_\infty} + \sum_{j_1=1}^{N} \frac{t^{a_1}_{j_1}}{z_{j_1}}\right)  \psi^{s_0, s_\infty}_0.
\end{align}
Note that the terms $t^{a_j}_{\infty}/z_{\infty}$ do not contribute
in the limit $z_\infty \to \infty$. On the unit circle, we have
\begin{align}
  \label{eq:Jpsi-zero-inf-from-FT-spin-operators}
  \Jpsi^{s_0, s_\infty}  &= \It{a_k}{-1} \dots \It{a_1}{-1} \psi^{s_0, s_\infty}_0
\end{align}
in terms of Fourier transformed spin operators.

\subsection{Odd Number of Spins}
\label{SEC:odd-number-of-spins}

A correlation function of vertex operators $\Braket{\Vi{1} \dots \Vi{N}}$ is only non-zero
if the sum $\sum_{i=1}^{N} s_i $ vanishes (cf. Eq. \eqref{eq:vertex-op-correlator}).
This property, the \emph{charge neutrality condition},
implies that the vertex operators in the correlator need to have a net charge of zero.
As a consequence, the number of vertex operators, and therefore the number
of spins, needs to be even.
We can, however, still consider a model with an odd number
of spins at the circle, by adding an extra vertex operator
that compensates the excess charge at the circle.
Inserting an additional vertex operator at $z=\infty$,
we obtain the state
\begin{align}
\label{eq:state-with-vertex-operator-at-infinity}
  &\psi^{s_\infty}_0 (s_1, \dots, s_N) \notag\\
  &\quad= \Braket{\V{s_\infty}{\infty} \Vi{1} \dots \Vi{N}} \notag \\
  &\quad\propto \delta_{\bar{\mathbf{s}}} (-1)^{(s_\infty+1)/2} \chi_{\mathbf{s}} \prod_{i < j}^{N} (z_i - z_j)^{s_i s_j / 2},
\end{align}
where $\delta_{\bar{\mathbf{s}}} = 1$ if $s_\infty + \sum_{i=1}^N s_i
   = 0$ and $\delta_{\bar{\mathbf{s}}} = 0$ otherwise.
The wave function $\psi^{s_\infty}_0$ has spin $1/2$,
\begin{align}
\label{eq:psi0add-has-spin-1-2} T^a T^a \psi^{s_\infty}_0 = \frac{3}{4} \psi^{s_\infty}_0, \quad \text{with } T^a = \sum_{i=1}^{N} t^a_i.
\end{align}
This is a consequence of $\psi^{s_\infty}_0$ being a singlet of the total spin including the point
$z=\infty$, $(t^a_\infty + T^a) \psi^{s_\infty}_0 = 0$.
As in the case of an even number of spins,
we define a tower of states by insertion of current operators,
\begin{align}
\label{eq:tower-above-psi0odd}
&\psi^{s_\infty}_{a_k \dots a_1}(s_1, \dots, s_N) \notag \\
&\quad= \Braket{\V{s_\infty}{\infty} \Vi{1} \dots \Vi{N} (\Ja{k} \dots \Ja{1})(0)}.
\end{align}
We obtain a second class of states by inserting the additional vertex operator at $z=0$
instead of $z=\infty$,
\begin{align}
\label{eq:alpha0odd-definition}
  &\psi^{s_0}_0(s_1, \dots, s_N) \notag\\
  &\quad= \Braket{\Vi{1} \dots \Vi{N} \V{s_0}{0}} \notag \\
  &\quad\propto \delta_{\bar{\mathbf{s}}} \chi_{\mathbf{s}} \prod_{i=1}^{N} z_i^{s_0 s_i/2} \prod_{i < j}^{N} (z_i - z_j)^{s_i s_j / 2},
\end{align}
and the corresponding tower
\begin{align}
\label{eq:alpha0odd-tower}
&\psi^{s_0}_{a_k, \dots, a_1}(s_1, \dots, s_N) \notag \\
&\quad = \Braket{\Vi{1} \dots \Vi{N} (\Ja{k} \dots \Ja{1} \Vs{s_0})(0)}.
\end{align}

On the unit circle, the two towers of states can be written as

\begin{align}
  \label{eq:odd-N-tower-in-terms-of-FT-spin-operators-psi0odd}
  \psi^{s_\infty}_{a_k, \dots, a_1}(s_1, \dots, s_N) &= \Ia{k} \dots \Ia{1} \psi^{s_\infty}_0
\end{align}
and
\begin{align}
  \label{eq:odd-N-tower-in-terms-of-FT-spin-operators-alpha0odd}
  \psi^{s_0}_{a_k, \dots, a_1}(s_1, \dots, s_N) &= \Ia{k} \dots \Ia{1} \psi^{s_0}_0,
\end{align}
respectively.

Comparing the wave functions of Eq. \eqref{eq:state-with-vertex-operator-at-infinity}
and Eq. \eqref{eq:alpha0odd-definition} with the case of an even number of spins
(Eq. \eqref{eq:vertex-op-correlator} and \eqref{eq:wave-function-alpha0}),
we conclude that $\psi^{s_\infty}_0$ is the analog of $\psi_0$ and $\psi^{s_0}_0$
of $\psi^{s_0 s_\infty}_0$, respectively.

\section{Parent Hamiltonian and Spectrum}
\label{SEC:Parent-Hamiltonian-and-Spectrum}
In this section, we introduce the parent Hamiltonian of $\psi_0$,
discuss its equivalence to the Haldane-Shastry model and
construct its spectrum from CFT current operators. Furthermore,
we relate our ansatz to the multiplets of the Yangian algebra
and the spinon construction of the CFT Hilbert space. From now
on we assume a uniform, one-dimensional lattice with
periodic boundary conditions,

\begin{align}
  \label{eq:fix-z-to-unit-circle}
  z_j = e^{2 \pi i j / N}
\end{align}
with $j \in \{1, \dots, N\}$.

\subsection{Parent Hamiltonian}
\label{SEC:Parent-Hamiltonian}
In earlier work on the Haldane-Shastry model \cite{Shastry1992},
an operator $\Co{a}{i}$ was constructed that annihilates the wave function $\Gpsi$,
$\Co{a}{i} \Gpsi = 0$. This operator was later obtained in a more general setting from
the $SU(2)_1$ WZW model using null vectors\cite{Nielsen2011}.
In terms of spin operators,
\begin{align}
\label{eq:definition-of-C}
\Co{a}{i} &= \frac{2}{3} \sum_{j(\neq i)} w_{ij} \left(t^a_j + i \vep_{abc} t^b_i t^c_j \right),
\end{align}
with
\begin{align}
\label{eq:definition-of-w}
w_{ij} &\equiv \frac{z_i + z_j}{z_i - z_j}.
\end{align}
We have used the notation $\sum_{i(\neq j)}$ for a sum over
all $i \in \{1, \dots, N\} \setminus \{j\}$.

This allows for the definition of a parent Hamiltonian $H$ of
$\Gpsi$ \cite{Nielsen2011},
\begin{align}
\label{eq:def-Hamiltonian}
H &= \frac{1}{4} \sum_{i=1}^N (\Co{a}{i})^\dagger \Co{a}{i}.
\end{align}
Note that $H$ is
positive semidefinite and $\Gpsi$ is an eigenstate of $H$ with zero energy.

It is known\cite{Nielsen2011} that $H$ is
closely related to the Haldane-Shastry Hamiltonian $H_{\text{HS}}$,
if the spins are uniformly distributed on the circle,
\begin{align}
\label{eq:H-and-HHS}
 H_{\text{HS}} &= \frac{1}{2} \sum_{i \neq j} \frac{t^a_i t^a_j}{\mathrm{sin}^2\left(\frac{(i-j) \pi}{N}\right)}
= H + \frac{N+1}{6} T^a T^a + E_0.
\end{align}
Here $T^a = \sum_{i=1}^{N} t^a$ is the total spin and $E_0 =
-(N^3+5N)/24$ the ground state energy of the Haldane-Shastry
Hamiltonian.  Note that $\Gpsi$ is annihilated by $H$ and also by
$T^aT^a$ in Eq. \eqref{eq:H-and-HHS} since it is a singlet. Therefore,
$\Gpsi$ is the ground state of the Haldane-Shastry Hamiltonian.

From now on, we will work with the Hamiltonian
\begin{align}
\label{eq:def-full-H} \Hf &= H + \frac{N+1}{6} T^a T^a,
\end{align}
dropping the constant $E_0$.

\subsection{Block Diagonal Form of the Hamiltonian}
\label{SEC:block-diagonal-form-of-the-Hamiltonian}
We now systematically construct excited states of $\Hf$ from
conformal correlation functions.  Specifically, we build the
excited states as linear combinations of the states $\Jpsi$
(cf. Eq. \eqref{eq:vertex-op-correlator} and \eqref{eq:result-Jpsi-from-Gpsi}).

The key to this construction is that the Hamiltonian does not couple
states with a fixed number of current operators to states with a
different number of current operators, i.e. the Hamiltonian is
block-diagonal in this basis.

Therefore, we can diagonalize the Hamiltonian in the subspaces of
states with a certain number of current operators.
It is not necessary to construct the Hamiltonian in the full Hilbert space of dimension $2^N$ in
order to find eigenstates beyond the ground state. Rather, we obtain
eigenstates by successively adding current operators and
diagonalizing the blocks.

Let us now show that $\Hf \Jpsi$ is a linear combination of states
obtained from $\Gpsi$ by insertion of $k$ current operators.

Recall that the insertion of $k$ current operator modes $J^{a_j}_{-1}$ $(j=1, \dots, k)$
into the correlation function of vertex operators is equivalent to the
successive application of Fourier transformed spin operators $\It{a_j}{-1}$ to
the ground state $\Gpsi$ (cf. Eq. \eqref{eq:def-I-spin-ops}).
Therefore, we have computed the commutator between $\Hf$ and $\It{a}{-1}$ by an
explicit expansion of the Hamiltonian
in terms of Fourier modes
$\It{a}{l}$ (cf. Section
\ref{SEC:Commutator-H-J} of the Appendix).  The result of this
calculation is
\begin{align}
\label{eq:commutator-Hf-with-I}
\Comm{\Hf}{\It{a}{-1}} &= (N-1) \It{a}{-1} + \sum_{i=1}^{N}\frac{3}{2} \frac{\Co{a}{i}}{z_i} + i \vep_{abc} \It{b}{-1} T^{c}.
\end{align}
From this we can already conclude that the energy of a state
with one current operator is $N-1$,
\begin{align}
\label{eq:energy-single-Jpsi}
\Hf \It{a}{-1} \Gpsi &= \Comm{\Hf}{\It{a}{-1}}\Gpsi = (N-1) \It{a}{-1} \Gpsi,
\end{align}
since $\Hf$, $\Co{a}{i}$, and $T^c$ annihilate the ground state $\Gpsi$.

We need to know how $\Co{a}{i}$ acts on $\psi_{a_k\dots a_1}$ to determine the energy
of states with more than one current operator.
As we show in Section \ref{SEC:decoupling-equation-for-Jpsi} of the Appendix,
\begin{align}
\label{eq:decoupling-equation-Jpsi}
  &\mathcal{C}_i^a \psi_{a_k\dots a_1} \notag \\
  &= \sum_{q=1}^k\frac{(K_{a_q}^a)_i}{z_i} \psi_{a_k\dots a_{q+1}a_{q-1}\dots a_1} \notag \\
  & \quad+(K_{b}^a)_i T^b\psi_{a_k\dots a_1} \notag \\
  & \quad +2 (K_{b}^a)_i \sum_{q=2}^k\sum_{n=0}^{q-1} \frac{i\vep_{ba_qc}}{z_i^{n+1}} \notag \\
  & \quad\quad \times \langle \PhiString (J_{-1}^{a_k}\dots J_{-1}^{a_{q+1}} J_n^{c}J_{-1}^{a_{q-1}} \dots J_{-1}^{a_1})(0)\rangle,
\end{align}
with
\begin{align}
  \label{eq:decoupling-equation-details}
  \PhiString &= \Vi{1} \dots \Vi{N}, \quad \text{and} \\
  (K^{a}_b)_{i} &= \frac{2}{3} (\delta_{ab} - i \vep_{abc} t^c_i).
\end{align}
Combining Eq. \eqref{eq:commutator-Hf-with-I} and
Eq. \eqref{eq:decoupling-equation-Jpsi}, we obtain
\begin{align}
\label{eq:derivation-H-Jpsi}
&\Hf \psi_{a_k\dots a_1} \notag \\
&= \sum_{r=1}^{k} \Ia{k} \dots \Ia{r+1} \Comm{\Hf}{\Ia{r}} \Ia{r-1} \dots \Ia{1} \psi_0 \notag \\
&= k (N-1) \psi_{a_k\dots a_1} + \sum_{2 \le q<r \le k} \sum_{n=0}^{q-1}  F_{a_k \dots a_1}^{qr,n}\notag \\
& \quad + \sum_{1 \le q<r \le k} \Big(2 \psi_{a_k\dots a_{r+1} a_q a_{r-1} \dots a_{q+1} a_{r} a_{q-1} \dots a_1} \notag \\
&\phantom{\quad + \sum_{q<r}^{k} \Big(} - 2 \delta_{a_r a_q} \psi_{a_k\dots a_{r+1} c a_{r-1} \dots a_{q+1} c a_{q-1} \dots a_1} \notag \\
&\phantom{\quad + \sum_{q<r}^{k} \Big(} + \psi_{a_k \dots a_{r+1} a_{q} a_{r} a_{r-1} \dots a_{q+1} a_{q-1} \dots a_{1}} \notag \\
&\phantom{\quad + \sum_{q<r}^{k} \Big(}  - \psi_{a_k \dots a_{r+1} a_{r} a_{q} a_{r-1} \dots a_{q+1} a_{q-1} \dots a_{1}} \notag \\
&\phantom{\quad + \sum_{q<r}^{k} \Big(} + 2 \delta_{N 2} \delta_{a_r a_q} \psi_{a_k \dots a_{r+1} a_{r-1} \dots a_{q+1} a_{q-1}\dots a_1}\Big)
\end{align}
with
\begin{align}
\label{eq:Fqrn}
&F_{a_k \dots a_1}^{qr,n} \notag \\
&= 2 \Braket{\PhiString ( \Ja{k} \dots \Ja{r+1} J^{a_q}_{-n-2} \Ja{r-1} \dots \Ja{q+1} \notag \\
&\quad\quad J^{a_r}_{n} \Ja{q-1} \dots \Ja{1})(0)} \notag \\
&\quad - 2 \delta_{a_r a_q} \Braket{\PhiString (\Ja{k} \dots \Ja{r+1} J^{c}_{-n-2} \Ja{r-1} \dots \Ja{q+1}  \notag \\
&\quad\quad  J^{c}_{n} \Ja{q-1} \dots \Ja{1})(0)} \notag \\
&\quad + 2 N \tilde{\delta}_{n+2} i \vep_{a_r a_q c} \Braket{\PhiString (\Ja{k} \dots \Ja{r+1} \Ja{r-1} \dots \Ja{q+1} \notag\\
&\quad\quad J^{c}_{n} \Ja{q-1} \dots \Ja{1}(0))(0)} .
\end{align}
In the last term, $\tilde{\delta}_{n+2} = 1$ if $(n+2)$ mod $N = 0$
and $\tilde{\delta}_{n+2} = 0$ otherwise.

Let us now argue that this expression contains $k$ current operators
of order $-1$.
There are two terms that are not yet explicitly written in the desired
form, namely the term proportional to $\delta_{N 2}$ and the term
abbreviated by $F_{a_k \dots a_1}^{qr,n}$.

The operators $J^{a_q}_{-n-2}$ (and $J^{c}_{-n-2}$, respectively), can
be written as a linear combination of $n+2$ current operators of order
$-1$ by repeated application of Eq. \eqref{eq:elim-higher-order-J}.
On the other hand, the operators $J^{a_r}_{n}$ (and $J^{c}_{n}$,
respectively), can be commuted to the right, using the current algebra
(cf. Eq. \eqref{eq:Kac-Moody-current-algebra})
\begin{align}
\label{eq:current-algebra-move-right}
\Comm{J^a_{n}}{J^b_{-1}} &= i \vep_{abc} J^{c}_{n-1} + \frac{1}{2} \delta_{ab} \delta_{n-1,0}.
\end{align}
The resulting terms either have $n$ current operators less or vanish
because $J^a_m \Ket{0} = 0$ for $m \ge 0$.

The total number of current operators in the first two terms of $F_{a_k \dots a_1}^{qr,n}$
is therefore
\begin{align}
\label{eq:total-number-of-current-operators}
k - 2 + \underbrace{n + 2}_{J^{a_q}_{-n-2}} \underbrace{-n}_{\vphantom{J^{a_q}_{-n-2}} J^{a_r}_{n}} = k.
\end{align}
If $k \ge N$, there can be a contribution from the third term in $F_{a_k \dots a_1}^{qr,n}$,
namely if $n+2 = m N$ for $m \in \{1, 2, \dots\}$.
This term can be written in terms of $k - m N$ current operators.
Note, however, that the space of states with $k$ current operators
contains the space of states with $k - m N$ current operators. The reason
for this is that a current operator $J^a_{-1}$ corresponds to the
application of a Fourier transformed spin operator $u^a_{-1}$,
for which $u^a_{-1} = u^a_{-1 - m N}$.
This argument also applies to the term that is proportional to $\delta_{N 2}$.

Thus, the Hamiltonian is block diagonal in the states
with a fixed number of current operators.

Note that this observation does not follow from translational invariance
only.  Translational invariance implies that the Hamiltonian does
not mix states with different lattice momenta. Since each
current operator in $\Jpsi$ contributes a unit of $2 \pi / N$
to the momentum, it follows from translational
invariance that $\Hf\Jpsi$ is a linear combination of states with $k
\text{ mod } N$ current operators.
The above considerations moreover show that it is possible to write $\Hf\Jpsi$
as a linear combination of states with strictly $k$ modes $J^a_{-1}$. In
particular, it is not necessary, to include terms with a higher number
of current operators.
This allows us to block-diagonalize the Hamiltonian starting
with the smaller blocks, i.e. those with a small number of current operators.

\subsection{Eigenstates from Current Operators}
\label{SEC:Eigenstates-Analytically}
We have solved the eigenvalue equation of
Eq. \eqref{eq:derivation-H-Jpsi} for up to eight current operator
modes analytically.  Since the Hamiltonian is $SU(2)$ invariant, we
have decomposed the eigenstates into different spin sectors.
The momenta of the states are directly related to the number of
current operators, with each current operator changing the
momentum by $2 \pi / N$.
We summarize our results for up to four current operators in Table \ref{table:computed-eigenstates}.

\begin{table*}[htb]
  \caption{\label{table:computed-eigenstates}
Eigenstates of $\Hf$ in terms of states obtained by insertion of $k$ current operator modes for $k \le 4$. The momentum of the ground state is $p_0 = \pi$ if $N/2$ is odd
and $p_0=0$ if $N/2$ is even.}
\centering
\begin{tabular}{rllllll}
$k$ & State & Null for & Energy & Spin & Momentum & Number of states \\
\hline
$0$ & $\varphi^{(0)} = \psi_0$ &  & $0$ & $0$& $p_0$ & $1$ \\
$1$ & $\varphi^{(1)}_a = \psi_a$ &  & $N - 1$ & $1$ & $p_0 - 2 \pi/N$ & $3$  \\
$2$ & $\varphi^{(2)} = \sum_c \psi_{c c} - 3 \delta_{N 2} \psi_0$& $N \le 2$ & $2(N-3)$ & $0$ & $p_0 - 4 \pi/N$ & $1$  \\
$2$ & $\varphi^{(3)}_a = \sum_{c d} \vep_{acd} \psi_{c d}$ & $N \le 2$ & $2(N-3)$ & $1$& $p_0 - 4 \pi/N$ & $3$  \\
$3$ & $\varphi^{(4)} = \sum_{cde} \vep_{cde} \psi_{cde}$ & $N \le 4$ & $3(N-5)$ & $0$ & $p_0 - 6 \pi/N$ & $1$ \\
$3$ & $\varphi^{(5)}_a = \sum_c (2 \psi_{acc} - 3 \psi_{cac} +  \psi_{cca}) - 4 \delta_{N2} \psi_a$ & $N \le 4$ & $3(N-5)$ & $1$ & $p_0 - 6 \pi/N$ & $3$ \\
$3$ & $\varphi^{(6)}_a = \sum_c (\psi_{cca} - \psi_{acc}) + 2 \delta_{N2} \psi_a$ & $N \le 2$ & $3(N-3)$ & $1$ & $p_0 - 6 \pi/N$ & $3$ \\
$4$ & $\varphi^{(7)} = \sum_{c d} (5 \psi_{c d c d} - 3 \psi_{c d d c} - 2 \psi_{c c d d}) + 16 \delta_{N 4} \psi_0 + 12 \delta_{N2} \psi_0$ & $N \le 6$ & $4 (N - 7)$ & $0$ & $p_0 - 8 \pi / N$ & $1$\\
$4$ & $\varphi^{(8)}_{a} = \sum_{c d e} (4 \vep_{a c d} \psi_{c d e e} - 3 \vep_{a c d} \psi_{c e d e})$ & $N \le 6$ & $4 (N-7)$ & 1 & $p_0 -8 \pi / N$ & $3$ \\
$4$ & $\varphi^{(9)} = \sum_{c d} (\psi_{c d d c} - \psi_{c c d d}) - 12 \delta_{N 4} \psi_0 + 6 \delta_{N2} \psi_0$ & $N \le 4$ & $4 N - 18$ &  $0$ & $p_0 - 8 \pi / N$ & $1$ \\
$4$ & $\varphi^{(10)}_{a} = \sum_{c d e} (\vep_{a c d} \psi_{c d e e} + 3 \vep_{a c d} \psi_{c e d e})$ & $N \le 4$ & $4 N - 18$ & $1$ & $p_0 - 8 \pi / N$ & $3$ \\
$4$ & $\varphi^{(11)}_{ab} = \sum_{c} \left(\frac{1}{2}(\psi_{a b c c } + \psi_{b a c c }) - \frac{1}{3} \delta_{ab} \sum_d \psi_{ddcc} \right)$ & $N \le 2$ & $4 N - 10$ & $2$ & $p_0 - 8 \pi / N$ & $5$
\end{tabular}
\end{table*}

At level one we find a triplet (spin one), at level two a
singlet and a triplet, at level three we find a singlet and two
triplets with different energies. A spin two state appears at
level 4 as the symmetric traceless part of a state with two non-contracted
$SU(2)$-indices.

Note that the number of eigenstates is smaller than the number of
possible combinations we can build with $k$ current operators. The
reason is that some CFT states $J^{a_k}_{-1} \dots J^{a_1}_{-1}
\Ket{0}$ are null, such that the norm of the corresponding spin state
vanishes.  At level three, for example, there is the null state
$\sum_b (3 \psi_0^{bab} + 3 \psi_0^{bba} - 2 \psi_0^{abb})$.

The number of states that we find with a certain number of
current operators is in agreement with the characters of the $SU(2)_1$
algebra\cite{Nielsen2011}: At level $0$, $1$, $2$, $3$, and 4 we find
$1$, $3$, $4$, $7$, and $13$ states, respectively.
The size of the matrices that need to be diagonalized at a given
level in current operators thus corresponds to the characters of
$SU(2)_1$.

When considering the action of the Hamiltonian on states with $k$
current operators, there are two types of terms that depend on the
number of spins $N$, cf Eq. \eqref{eq:derivation-H-Jpsi}.  The first
one, $k (N-1) \Jpsi$, is already diagonal. All other terms only appear
if $k \ge N$. These terms are strictly upper triangular in the sense that
they can be written in terms of $k - m N$ current operators with $m >
0$. This upper triangular structure is preserved by a diagonalization
of all other terms. Therefore, only the diagonal term $k (N-1) \Jpsi$
contributes to the $N$-dependence of the energies.

We thus arrive at an $N$-independent representation of the energies by
subtracting the contribution $k (N-1)$ from the energies,
\begin{align}
  \label{eq:shifted-and-rescaled-energies} \tilde{E} &= \begin{cases}
\frac{E - (N-1) k}{k}, &k > 0,\\ 0, &k = 0.
               \end{cases}
\end{align}
(We have also rescaled the energies by $1/k$ for
convenience.)

\begin{figure*}[htb]
\centering
\includegraphics[width=.9\linewidth]{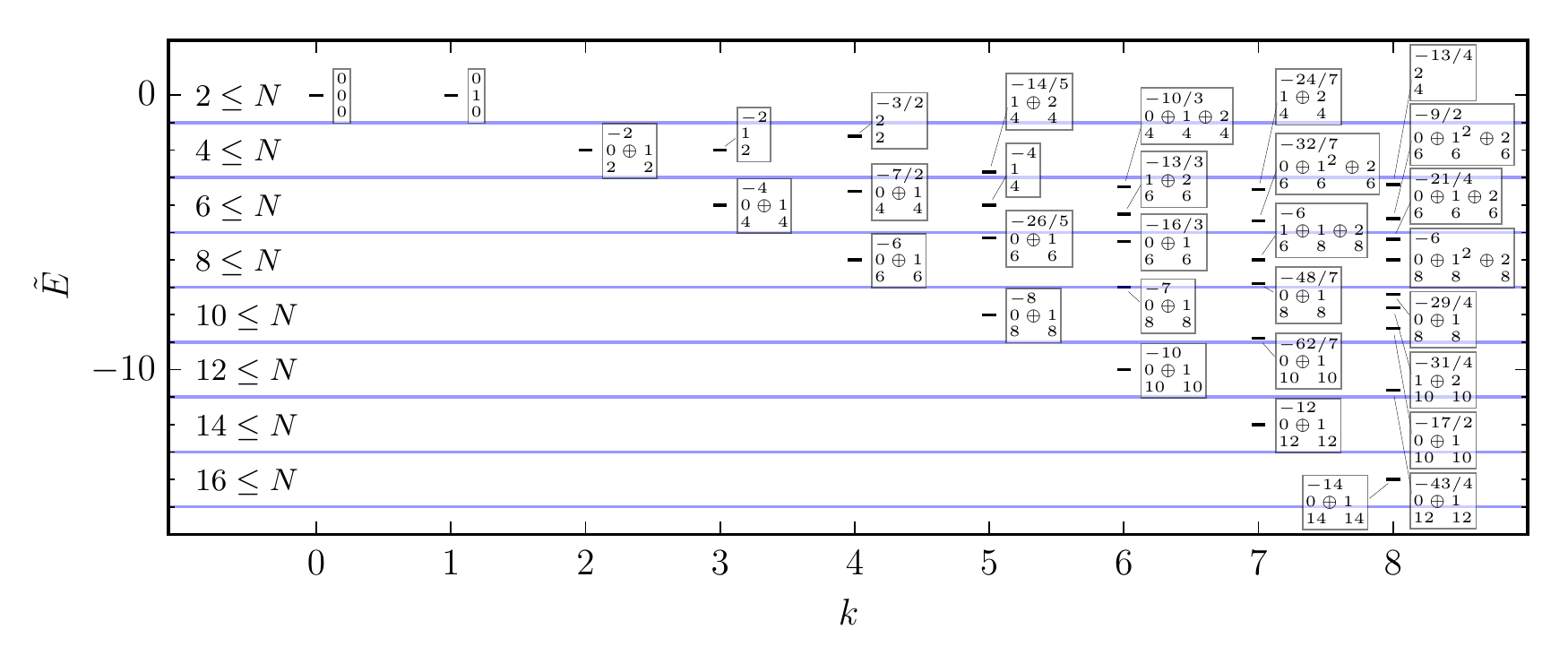}
\caption{\label{fig:analytical_spectrum} (Color online) Analytically
calculated energy levels and their spin content as a function of the
number of current operators $k$. $\tilde{E}$ is defined as $\tilde{E}
= (E - (N-1) k)/k$ for $k > 0$ and $\tilde{E} = E = 0$ for $k=0$. The
three rows in the boxes next to the energy level correspond to the
calculated values of (1) the energy $\tilde{E}$ (2) the spin content
(3) a value for the number of spins $N'$ so that the corresponding
state is null for all systems with $N \le N'$ spins.  The inequalities
indicate for which $N$ the condition $E \ge 0$ is satisfied. For a
given number of spins $N$, all eigenvalues that lie in bands where the
inequality is not satisfied correspond to null states. By applying
additional current operators to these null states, we could
identify further null states that do not violate the energy condition.}
\end{figure*}

The shifted and rescaled energies $\tilde{E}$ as well as the spin
content of the corresponding eigenspaces are plotted for up to
eight current operators in Fig. \ref{fig:analytical_spectrum}.

Note that for a given $N$, some of these energies do not
occur because the corresponding states have zero norm. These null
states appear dependent on $N$ and in addition to null states
identified at the CFT level. The occurrence of additional null states
reflects the fact that the CFT Hilbert space is infinite, while the
spin system's Hilbert space is finite.

A necessary condition for a particular state to be non-null is given
by $E \ge 0$. The region of energies, where $E \ge 0$ is satisfied is
indicated by the bands in
Fig. \ref{fig:analytical_spectrum}. Depending on $N$, we find certain
states that violate this condition and are therefore null. These
states then lead to further null states at higher levels through the
application of additional current operators. We find that if a state
corresponding to a certain energy is null for a number of spins $N'$,
it is also null for $N$ spins with $N \le N'$. For each energy level,
the highest $N'$ that we found exploiting the energy condition $E \ge
0$ is given in Fig.~\ref{fig:analytical_spectrum} below the spin
content.

Let us give an example for the notation used in
Fig. \ref{fig:analytical_spectrum}. At $k=7$ and $\tilde{E}=-6$ we
find two spin $1$ states and one spin $2$ state.  One of the spin $1$
states is null for all $N \le 6$, the other spin $1$ state and the
spin $2$ state are null for all $N \le 8$. This multiplet is shown in
Fig. \ref{fig:analytical_spectrum} as

\begin{equation}
  \label{eq:example-of-multiplet}
  \boxed{
    \begin{matrix}-6 & & & &\\1 &\oplus &1 &\oplus &2 \\ 6 && 8 && 8
    \end{matrix}
  }.
\end{equation}

Note that the violation of the energy condition $E \ge 0$ is
sufficient for a state to be null. A complete separation of null
states requires the computation of inner products between our ansatz
states at the level of the spin system. As shown in
Ref. \onlinecite{Nielsen2011}, the spin correlation functions in $\psi_0$
can be computed by solving linear algebraic equations. Using
these correlation functions, one could compute the norms of the states
$\Jpsi$ numerically, even for large system sizes.

\subsection{Construction of Yangian Highest Weight States and Comparison to Spinon Basis}
The spectrum of the Haldane-Shastry model has previously been
constructed by exploiting a hidden symmetry of the Hamiltonian, which
is generated by the rapidity operator $\Lambda^a$\cite{Haldane1992},
\begin{align}
  \label{eq:definition-of-Yangian}
  \Lambda^a &= \frac{i}{2} \sum_{i \neq j} w_{ij} \vep_{abc} t^b_i t^c_j = \frac{i}{2} \sum_{i \neq j} w_{ij} (\vec{t}_i \times \vec{t}_j)^a.
\end{align}
Using $\sum_{j(\neq i)} w_{ij} = 0$, it follows that
\begin{align}
  \label{eq:rapidity-in-terms-of-c}
  \Lambda^a = \frac{3}{4} \sum_{i=1}^N \Co{a}{i}.
\end{align}
The rapidity $\Lambda^a$ and the total spin $T^a$ form a basis of the
Yangian algebra.  They both commute with the Haldane-Shastry
Hamiltonian, but the rapidity $\Lambda^a$ does not commute with the
total spin operator $T^a T^a$.  This is the reason for the degenerate
energy levels formed by multiplets with different total
spin\cite{Haldane1992}.

The key to the construction of eigenstates in this approach is the
notion of a Yangian highest weight state $h$, which is annihilated by
$\Lambda^{+} = \Lambda^{x} + i \Lambda^{y}$ and $T^{+} = T^{x} + i T^{y}$.
A multiplet of states with the same energy is then given by application
of powers of $\Lambda^{-} = \Lambda^{x} - i \Lambda^{y}$ to $h$.

In order to relate our method to this approach, we have computed
the action of $\Lambda^a$ on $\Jpsi$ using the decoupling equation derived
in section \ref{SEC:decoupling-equation-for-Jpsi} of the Appendix.
We find
\begin{align}
  \label{eq:action-of-Yangian}
2 \Lambda^a \Jpsi &= \sum_{q=1}^k (N-1) i \vep_{a a_q c} \psi_{a_k \dots a_{q+1} c a_{q-1} \dots a_1} \notag \\
              &\quad + \sum_{q=1}^k i \vep_{a_q a c} \psi_{c a_{k} \dots a_{q+1} a_{q-1} \dots a_{1}} \notag \\
              &\quad + \sum_{q=2}^k \sum_{n=0}^{q-1} G^{q, n}_{a_k \dots a_q},
\end{align}
with
\begin{align}
  \label{eq:G-symbol-defintion}
  &G^{q, n}_{a_k \dots a_q} = \notag \\
  &2 \langle \PhiString (J^{a_q}_{-n-1} \Ja{k} \dots \Ja{q+1} J^a_n \Ja{q-1} \dots \Ja{1})(0) \rangle \notag \\
  &- 2 \delta_{a_q a} \langle \PhiString (J^c_{-n-1} \Ja{k} \dots \Ja{q+1} J^{c}_n \Ja{q-1} \dots \Ja{1})(0) \rangle\notag\\
  &+2 N \tilde{\delta}_{n+1} i \vep_{a a_q c}  \langle \PhiString (\Ja{k} \dots \Ja{q+1} \notag\\
  &\quad\quad J^{c}_n \Ja{q-1} \dots \Ja{1})(0) \rangle.
\end{align}
Furthermore, we have (cf. Eq. \eqref{eq:T-Jpsi-in-epsilon} in Appendix \ref{SEC:decoupling-equation-for-Jpsi})
\begin{align}
  \label{eq:action-of-T-on-Jstate}
  T^a \Jpsi &= i \sum_{q=1}^N \vep_{a a_qc} \psi_{a_k \dots a_{q+1} c a_{q-1}\dots q_1}.
\end{align}

This shows that the Yangian, like the Hamiltonian, leaves the
subspaces of states with a fixed number of current operators
invariant. It is thus possible to write the highest weight states of
the Yangian algebra in terms of the states $\Jpsi$. We have computed
the highest weight states for up to four current operators and
expanded the result in the eigenstates listed in Table
\ref{table:computed-eigenstates}. These states thus correspond to the
eigenstates constructed by Haldane in Ref.~\onlinecite{Haldane1991},
which have a polynomial form in the particle basis and were identified
as the highest weight states of the Yangian algebra in
Ref.~\onlinecite{Haldane1992}. We summarize our results in Table
\ref{table:highest-weight-states}.

\begin{table}[htb]
  \caption{\label{table:highest-weight-states}
  Highest weight states of the Yangian algebra in terms of the eigenstates of the Hamiltonian
  given in Table \ref{table:computed-eigenstates}.}
\centering
\begin{tabular}{rlllll}
$k$ & State &  Energy & Spin \\
\hline
$0$ & $\varphi^{0}$ & $0$ & $0$ \\
$1$ & $\varphi^{(1)}_x + i \varphi^{(1)}_y$ & $N-1$ & $1$ \\
$2$ & $\varphi^{(3)}_x + i \varphi^{(3)}_y$ &  $2(N-3)$ & $1$ \\
$3$ & $\varphi^{(5)}_x + i \varphi^{(5)}_y $ & $3(N-5)$ & $1$ \\
$3$ & $\varphi^{(6)}_x + i \varphi^{(6)}_y$ & $3(N-3)$ & $1$ \\
$4$ & $\varphi^{(8)}_x + i \varphi^{(8)}_y$ & $4 (N-7)$ & $1$ \\
$4$ & $\varphi^{(10)}_x + i \varphi^{(10)}_y$ & $4N - 18$ & $1$ \\
$4$ & $\varphi^{(11)}_{zz} + 2 \varphi^{(11)}_{xx} + 2i \varphi^{(11)}_{xy}$ & $4 N - 10$ & $2$
\end{tabular}
\end{table}
Note that our ansatz is manifestly $SU(2)$ invariant, whereas
the construction in terms of highest weight states is not.
The states $\Jpsi$ have a simple form in the sense that
they are created from the ground state by successive application
of Fourier transformed spin operators (cf. Eq. \eqref{eq:def-I-spin-ops}).

We have compared the highest weight states of the spin system to the highest
weight states of the Yangian algebra in the CFT, which is spanned
by\cite{Haldane1992} $J^a_{0}$ and
\begin{align}
  \label{eq:CFT-lambda}
  Q^a = \frac{i}{2} \sum_{m=1}^{\infty} \vep_{abc} J^b_{-m} J^c_{m}.
\end{align}
We find that the highest weight states that we computed in the spin system
are precisely the highest weight states of $J^a_0$ and $Q^a$, when
seen as states of the CFT under the correspondence of
Table \ref{table:structure-of-excited-states}.

Finally, let us relate our ansatz to the spinon basis of the
CFT Hilbert space \cite{Bouwknegt1994}. In this approach,
states are not constructed by application of modes of the affine current $J^a_{-n}$
to the CFT vacuum, but by application of modes $\phi_{s, -m}$ of the primary field $\phi_s(z)$.
These modes are defined with respect to the Laurent expansion \cite{Bouwknegt1994}
\begin{align}
  \label{eq:Laurent-expansion-of-spinon-modes}
  \phi_s(z) &= \sum_{m} z^{m + \frac{q}{2}} \phi_{s, -m-\frac{1}{4}-\frac{q}{2}}.
\end{align}
Here $q \in \{0, 1\}$ corresponds to the sector of the CFT Hilbert space on which $\phi_s(z)$
is acting: $q=0$ for a state built from an even number of spinon modes and $q=1$ for a
state with an odd number of spinon modes. The authors
of Ref.~\onlinecite{Bouwknegt1994} derived generalized commutation relations
for the modes of $J^a(z)$ and $\phi_s(z)$, which can be used
to rewrite states built from current operator modes in terms of
spinon modes $\phi_{s, -m}$. At level $k=1$, for example,
the state $\Braket{\Vi{1} \dots \Vi{N} J^a_{-1}(0)}$ corresponds to
\begin{align}
  \label{eq:J-lowest-level-to-spinon-modes}
  \sum_{s s'} (t^a)_{s s'} \Braket{\Vi{1} \dots \Vi{N}  \left(\phi_{-s, -\frac{3}{4}} \phi_{s', -\frac{1}{4}}\right)(0)}.
\end{align}
A general state with $k$ current operators of order $-1$ will be a linear
combination of terms with spinon modes $\phi_{\alpha_l, -m_l} \dots \phi_{\alpha_1, -m_1}$ with $k = \sum_{i=1}^{l} m_i$.

\subsection{Complete Spectrum}
\label{SEC:Numerics}
In the previous subsections, we have analytically diagonalized the
Hamiltonian in the states $\Jpsi$ for $k \le 8$ and $N$
even. For a given $k$, the size of the matrices that need to be
diagonalized and therefore the complexity of the analytical
calculation does not depend on the number of spins $N$. However, it
becomes increasingly difficult for larger $k$.  In order to test if our
method yields all excited states or just a subset thereof, we have
thus constructed the states $\Jpsi$ numerically for small $N$ and
performed a numerical diagonalization of the Hamiltonian in that
basis.  Our numerical calculations confirm that the complete
spectrum is indeed obtained from our ansatz states.

In an analogous way to $\Gpsi$, we studied the tower of states
obtained from $\psi^{s_0, s_\infty}_0$ by insertion of current
operators (cf. Section \ref{SEC:definition-psi-zero-inf}).  We find
numerically that the complete Hilbert space is generated, starting
from $\psi^{s_0s_\infty}_0$ with $s_0, s_\infty \in \{-1, 1\}$ and
successively inserting current operator modes.  As for $\psi_0$, the
Hamiltonian is block diagonal in the states with a fixed number of
current operators.

Furthermore, we constructed the spectrum numerically for the case of
an odd number of spins (cf. Section \ref{SEC:odd-number-of-spins}).
We find that the Hamiltonian is block diagonal in the states
$\Jpsi^{s_0}$ and $\Jpsi^{s_\infty}$ and that the complete Hilbert
space can be constructed from states of this form.

\begin{figure}[htb]
\centering
\includegraphics[width=.9\linewidth]{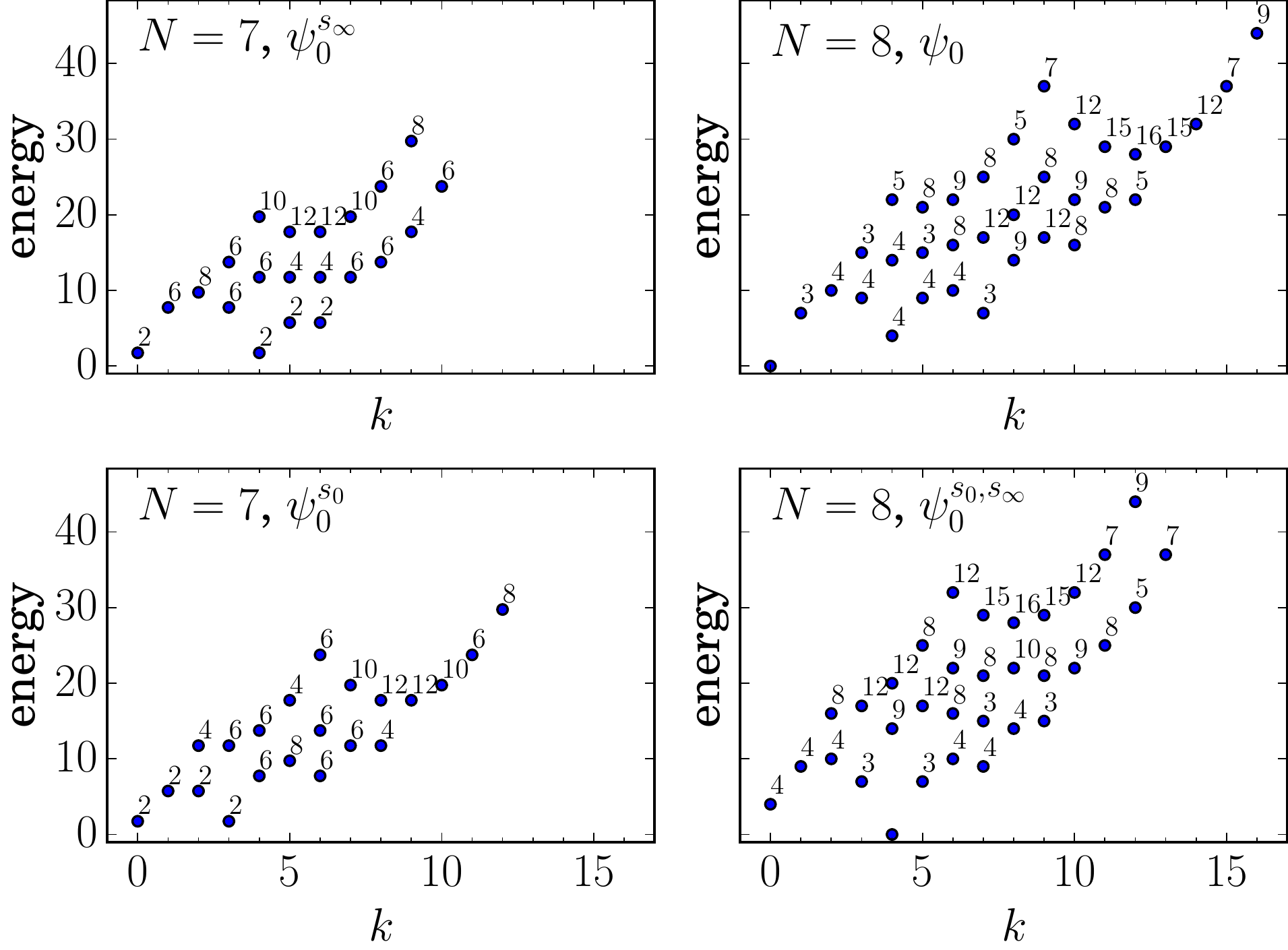}
\caption{\label{fig:numerical-spectra}
Numerically calculated spectra for the odd-$N$ spin chain ($N=7$, left
panels) and the even-$N$ spin chain ($N=8$, right panels).  The upper
panels show data obtained by applying current operators to the states
$\psi^{s_\infty}_0$ and $\psi_0$, respectively.  In the lower panels,
the current operators were applied to the states $\psi^{s_0}_0$ and
$\psi^{s_0 s_\infty}_0$, respectively.  The horizontal axis shows the
number of current operators $k$ that are present in the corresponding
states. The numbers above the levels indicate their degeneracy.
Notice that the spectrum is the same in the upper and lower plots, but
a given state does not necessarily appear at the same number of
current operators.}
\end{figure}

The numerically
obtained spectra are shown for $N=7$ and $N=8$ spins in
Fig. \ref{fig:numerical-spectra}.

We observe a tendency that states with a higher
number of current operators have a larger energy.
This means that by inserting few current operators, we get access to
low-lying excited states.  Note, however, that this
relation does not hold in a strict sense; it may happen that, by
inserting additional current operators, we obtain states with a lower
energy.

We observe that the spectra shown in the upper and in the lower panels
of Fig. \ref{fig:numerical-spectra} are the same. This
means that for both $N$ even and $N$ odd, the complete spectrum
is obtained for either of the two classes of states.

The top right panel of Fig. \ref{fig:numerical-spectra} corresponds to
our analytical results shown in Fig. \ref{fig:analytical_spectrum}.
Note that not all states of Fig. \ref{fig:analytical_spectrum} appear
in the numerically calculated spectrum, because certain states of
Fig. \ref{fig:analytical_spectrum} are null for $N=8$. This is the
case for the all states with $E < 0$, which appear in the bands
below $\tilde{E} = -7$ in Fig. \ref{fig:analytical_spectrum}. The
remaining null states have the values
\begin{center}
\begin{tabular}{llll}
$k$ & $\tilde{E}$ & $E$ & Spin\\
\hline
$6$ & $-7$ & $0$ & $0 \oplus 1$\\
$7$ & $-\frac{48}{7}$ & $1$ & $0 \oplus 1$\\
$7$ & $-6$ & $7$ & $1 \oplus 2$\\
$8$ & $-6$ & $8$ & $0 \oplus 1^2 \oplus 2$\\
\end{tabular}
\end{center}
and can be obtained from null states violating the energy
condition by the insertion of additional current operators.

\begin{figure}[htb]
  \centering
  \includegraphics[width=.9\linewidth]{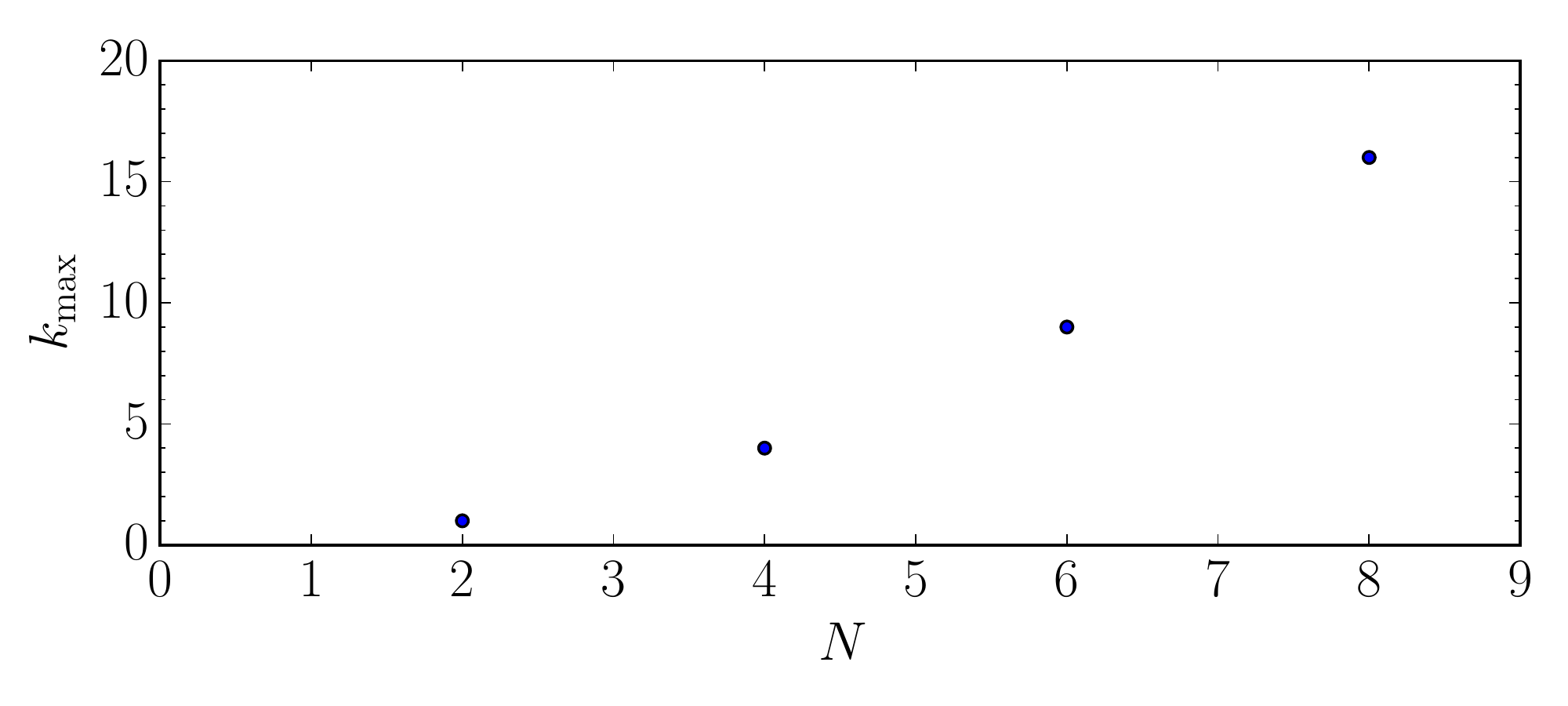}
  \caption{\label{fig:kmax-vs-N-for-N-even-and-psi0}
    Maximal number of current operator insertions $k_{\mathrm{max}}$
    of order $-1$ needed to obtain the spectrum of the Hamiltonian
    dependent on the number of spins $N$.  The data is consistent with
    $k_{max} = (N/2)^2$.}
\end{figure}

We have numerically computed the maximal number of current operator
insertions of order $-1$ needed to generate the complete spectrum as a
function of $N$. The results are shown for the tower of states built
on $\psi_0$ ($N$ even) in
Fig. \ref{fig:kmax-vs-N-for-N-even-and-psi0}. In this case, our
numerical data suggests that $(N/2)^2$ current operators are
sufficient to obtain the complete spectrum.

Finally, let us comment on how the first
excited states of the Haldane-Shastry model for $N$ even
are related to the states $\Jpsi$.
Except for $N=2$, these cannot be given by the states at
level one, $\psi_{a_1}$, which have $E=N-1$. The reason is that the states
$\psi^{s_0, s_\infty}_0$ are eigenstates with a lower energy of $E=N/2$. This follows
from
\begin{align}
  \label{eq:decoupling-psi-0inf}
  \mathcal{C}^a_i \psi^{s_0, s_\infty}_0 &= (K^a_b)_i (t^b_{\infty} - t^b_{0}) \psi^{s_0, s_\infty}_0
\end{align}
and
\begin{align}
  \label{eq:cdagger-on-circle}
  (\mathcal{C}^a_i)^{\dagger} = \mathcal{C}^a_i  - \frac{4}{3} \sum_{j (\neq i)} w_{ij} t^a_j.
\end{align}
Note that the latter equality is only valid on the circle with $z_j = e^{2 \pi i j / N}$.  Our
numerical calculations indicate that the states $\psi^{s_0,
s_{\infty}}_0$ are indeed the first excited states. Furthermore, we find a multiplet
with spin content $0 \oplus 1$ and energy $N/2$ at $k=N/2$ among the states constructed
from $\Jpsi$, both in the numerical and in the analytical calculations.  These multiplets
appear at $\tilde{E} = -2 k + 2$ in the analytically computed
spectrum shown in Fig. \ref{fig:analytical_spectrum}.
Noting that the states $\psi^{s_0, s_\infty}_0$ can be decomposed into a singlet
and a triplet, this suggests that the first excited states
are given by linear combinations of states $\Jpsi$ with $k=N/2$
current operator modes.

\section{Conclusion}
\label{SEC:Conclusion}
We studied a model of a quantum spin chain that is constructed
from the $SU(2)_1$ WZW model and is equivalent to the Haldane-Shastry model.
We have shown that it is possible to construct excited states
of the spin system from CFT current operators in a
way, which directly reflects the structure of
the CFT spectrum.

In the approach pursued in this work,
correlation functions of operators in the CFT
are interpreted as wave functions of spin states.
Based on earlier work,
where the correlation function of \emph{primary} operators
were interpreted as the ground state wave function, we provided a method
of constructing excited states by inserting \emph{descendant} fields.

In the case of an even number of spins $N$, we have diagonalized the
Hamiltonian analytically for $k \le 8$ inserted current operators.
Depending on $N$, we identified certain states that are null but
correspond to non-null CFT states. These additional null states occur
due to the finite size of the spin system's Hilbert space. Our method
of detecting them is based on a sufficient criterion for a state to be
null.  In order to test if a given state is not null, one could use
the known algebraic equations for spin correlation
functions\cite{Nielsen2011} in $\psi_0$ to compute the needed inner
products numerically, even for large system sizes.

We have given numerical evidence that this method yields the complete
spectrum for a given number of spins $N$, independent of which primary
state we build the tower states from. Furthermore, we have shown
numerically that a similar construction can be made for an odd number
of spins.

Our manifestly $SU(2)$ invariant ansatz is
compatible with the construction of eigenstates
of the Haldane-Shastry model as multiplets of the Yangian algebra
in the sense that the Yangian operator does not change the number
of current operators. This allowed us to explicitly relate
the excited states constructed from current operators to
the highest weight states of the Yangian operator. Furthermore, we
have argued that our ansatz wave functions with $k$ current operator
modes of order $-1$ can be rewritten in terms of a wave function
with spinon modes whose mode numbers sum up to $-k$.

In the case of the $SU(2)_1$ WZW model, which we studied here, the
resulting spin system is equivalent to the Haldane-Shastry model.
Thus our method provides an alternative way of constructing the
excited states of the Haldane-Shastry model, which emphasizes its
close relation to the underlying CFT. We expect that this method could
be generalized to the $SU(n)_1$ WZW model, which is related to the
$SU(n)$ Haldane-Shastry spin chain \cite{Schoutens1994}.  Another
generalization of our construction could be possible within the
Laughlin lattice models with filling fraction $1/q$
\cite{Tu2014b}. Even though these systems do not have a Yangian
symmetry for $q > 2$, part of the spectrum is described by integer
eigenvalues. We have carried out exemplary numerical calculations for
$q=3$ which indicate that at least some of the excited states can be
obtained by insertion of current operators.

It is crucial for the construction used above that the system
is one-dimensional with periodic boundary conditions.  This can be
seen, for example, noting that the ansatz $\Jpsi$ is a decomposition
into momentum space eigenstates.  However, the states obtained by
insertion of current operators might still describe low energy
eigenstates of an interesting, two-dimensional system with a different
Hamiltonian.  We plan to investigate the properties of these states in
two dimensions in future work.

\begin{acknowledgments}
 We are grateful to J. Ignacio Cirac for many fruitful discussions.
This work has been supported by the EU project SIQS, FIS2012-33642,
QUITEMAD (CAM), and the Severo Ochoa Program.  Figures were created
using \emph{matplotlib} \cite{Hunter2007} and \emph{MayaVi}
\cite{Ramachandran2011}.
\end{acknowledgments}

\appendix
\section{Commutator Between $\mathcal{H}$ and $\It{a}{-1}$}
\label{SEC:Commutator-H-J}
In this section we derive an expression for the commutator
$\Comm{\Hf}{\It{a}{-1}}$.  This commutator was used in
Section~\ref{SEC:block-diagonal-form-of-the-Hamiltonian} to determine
the action of $\Hf$ on a state $\Jpsi$.  The explicit form of the
Hamiltonian in terms of spin operators is \cite{Nielsen2011}
\begin{align}
  \label{eq:H-in-spin-operators}
  \Hf &= -\frac{1}{6} \sum_{k,j}\sum_{i\neq(k, j)}w_{ij} w_{ik} t^b_j t^b_k \notag \\
  &\quad - \frac{1}{6} \sum_{i\neq j} w_{ij}^2 t^b_i t^b_j + \frac{1}{6} (N+1) T^b T^b.
\end{align}
We use the Fourier transforms
\begin{align}
  \label{eq:ft-w-symbols}
\sum_{i \neq j}
\frac{w_{ij}}{z_i^k z_j^l} &=
\begin{cases}
0, &\text{if } k = 0
\\ (2k-N) \tilde{\delta}_{k+l} N,&\text{if } k = 1, 2, \dots, N-1 \end{cases} \notag
\\ \sum_{i \neq j} \frac{w_{ij}^2}{z_i^k z_j^l} &=
\begin{cases}
(N - N^2/3 - 2/3) \tilde{\delta}_{l} N, \\
\quad\text{if } k = 0\\
(N^2/6 - 2 (k - N/2)^2 - 2/3) \tilde{\delta}_{k+l} N, \\
\quad\text{if } k = 1, 2, \dots, N-1.
\end{cases}
\end{align}
to rewrite $\Hf$ in terms of $\It{a}{k}$.
Here $\tilde{\delta}_{m} = 1$ if $m \text{ mod } N = 0$
and $\tilde{\delta}_{m} = 0$ otherwise.
A derivation of these formulas is given in
Section \ref{SEC:FT-of-w} of the Appendix, see
also Ref. ~\onlinecite{Bernevig2001}, where these
Fourier sums were evaluated using contour integrals.

We find
\begin{align}
  \label{eq:H-in-I-operators}
  \Hf &= \sum_{k=0}^{N-1} \frac{1 + 2 N^2 - 9 N k + 9 k^2}{9N} \It{b}{-k}\It{b}{k}.
\end{align}
Starting from this expansion, we next compute the
commutator between $\Hf$ and $\It{a}{-1}$. Using
\begin{align}
  \label{eq:commutator-algebra-of-FT-spin-operators}
  \Comm{\It{a}{m}}{\It{b}{n}} &= i \vep_{abc} \It{c}{n+m},
\end{align}
which directly follows from the commutator algebra of
the spin operators $t^a_i$, we obtain
\begin{align}
  \label{eq:H-comm-ft-in-terms-of-FT-spin-operators}
  \Comm{\Hf}{\It{a}{-1}} &= 2 i \vep_{abc} \It{b}{-1} T^{c}
                           - \frac{2 i}{N} \vep_{abc} \sum_{m=1}^{N-1} m \It{b}{-m} \It{c}{m-1} \notag \\
  &\quad + (3-N) \It{a}{-1}.
\end{align}
It is possible to rewrite the sum over Fourier transformed spin operators
in terms of the operator $\Co{a}{i}$, which was defined in Eq. \eqref{eq:definition-of-C}.
To this end, we computed the Fourier expansion of $\Co{a}{i}$. The result is
\begin{align}
  \label{eq:Fourier-expansion-of-c}
  \frac{3}{2} \sum_{j=1}^{N} \frac{\Co{a}{j}}{z^p_{j}} &= i \vep_{abc} \It{b}{-p} T^c + (2 p + 2 - 2 N) \It{a}{-p} \notag \\
  &\quad - \frac{2 i}{N} \vep_{abc} \sum_{m=1}^{N-1} m \It{b}{-m} \It{c}{m-p},
\end{align}
for $p \in \{1, \dots, N-1\}$. Note that for $p=1$ we find
the same sum over Fourier transformed spin operators that occurs
in the above expansion of the Hamiltonian (Eq. \eqref{eq:H-comm-ft-in-terms-of-FT-spin-operators}).
Inserting the $p=1$ Fourier mode of $C^a_i$, we arrive at
\begin{align}
  \label{eq:final-result-commutator-H-c}
  \Comm{\Hf}{\It{a}{-1}} &= (N-1) \It{a}{-1} + i \vep_{abc} \It{b}{-1} T^c + \frac{3}{2} \sum_{i=1}^N \frac{\Co{a}{i}}{z_i}.
\end{align}

\section{Fourier Transform of $w$-Functions}
\label{SEC:FT-of-w}
The Fourier transforms of $w_{ij} = (z_i + z_j)/(z_i - z_j)$ and
$w_{ij}^2$ can both be reduced to one sum by translational
invariance,
\begin{align}
\label{eq:derivation-FTw-tranlational-invariance}
  \sum_{i \neq j} \frac{w_{ij}}{z_i^k z_j^l} &= N \tilde{\delta}_{k+l} \sum_{i (\neq N)} \frac{w_{iN}}{z_i^k},\\
  \allowdisplaybreaks
  \sum_{i \neq j} \frac{w_{ij}^2}{z_i^k z_j^l} &= N \tilde{\delta}_{k+l} \sum_{i (\neq N)} \frac{w_{iN}^2}{z_i^k}.
\end{align}

In order to evaluate the remaining sums, it is useful to compute
the Fourier sums
\begin{align}
\label{eq:derivation-FTw-kn}
\sum_{k=0}^{N-1} k^n z_k^j
\end{align}
for $n=1$ and $n=2$. For $j=0$, we have
\begin{align}
\label{eq:derivation-FTw-k0-ft}
  \sum_{k=0}^{N-1} k &= \frac{N (N-1)}{2},\\
  \sum_{k=0}^{N-1} k^2 &= \frac{N (N-1) (2N-1)}{6}.
\end{align}
For $j = 1, 2, \dots, N-1$, we use the generating function
\begin{align}
\label{eq:derivation-FTw-generating-function}
f(\omega) &= \sum_{k=0}^{N-1} e^{i \omega k} = \frac{1-e^{i \omega N}}{1 - e^{i \omega}}, \\
\sum_{k=0}^{N-1} k^n z_k^j &= \left(\frac{\ud}{i \ud \omega}\right)^n f(\omega) \Bigr\rvert_{\omega = 2 \pi j/N}. \notag
\end{align}
Taking the first and the second derivative, we find
\begin{align}
\label{eq:drivation-FTw-derivatives}
  \sum_{k=0}^{N-1} k z_k^j &= \frac{N}{2} (w_{jN} - 1), \notag \\
  \allowdisplaybreaks
  \sum_{k=0}^{N-1} k^2 z_k^j &= \frac{N}{2} (1-N+N w_{jN} - w_{jN}^2).
\end{align}
Taking the inverse Fourier transforms of these equations and solving for $\sum_{j (\neq N)} w_{jN}/z_j^k$ and
$\sum_{j (\neq N)} w_{jN}^2/z_j^k$, we arrive at Eq. \eqref{eq:ft-w-symbols}.

\section{Decoupling Equation for States with Current Operators}
\label{SEC:decoupling-equation-for-Jpsi}
The starting point for the derivation of the action of $\Co{a}{i}$
on a state
\begin{align}
\label{eq:appendix-decoupling-Jpsi-repeated-def}
&\Jpsi(s_1, \dots, s_N)\notag\\
&= \Braket{\Vi{1} \dots \Vi{N} (J^{a_k}_{-1} \dots J^{a_1}_{-1})(0)}
\end{align}
is the null operator \cite{Nielsen2011}
\begin{align}
\label{eq:null-operator}
  (K^{a}_{b})_i &(J^b_{-1} \varphi_{s_i})(z_i),\quad \text{with}\\
  (K^{a}_{b})_i &= \frac{2}{3} (\delta_{ab} - i \vep_{abc} t^c_i). \notag
\end{align}
When inserting the null operator into a correlation function
of primary fields, the resulting correlator vanishes.
We therefore have
\begin{widetext}
\begin{align}
0&=(K_{b}^a)_i\langle \phi_{s_1}(z_1)\ldots
   (J_{-1}^b\phi_{s_i})(z_i) \ldots\phi_{s_N}(z_N) (J^{a_k}_{-1}J^{a_{k-1}}_{-1}\ldots J^{a_1}_{-1})(0)\rangle \notag \allowdisplaybreaks \\
 &= (K_{b}^a)_i \oint_0\frac{\ud w_k}{2 \pi i w_k}\cdots\oint_0\frac{\ud w_1}{2 \pi i w_1}\langle
   \phi_{s_1}(z_1)\ldots (J_{-1}^b\phi_{s_i})(z_i)
   \ldots\phi_{s_N}(z_N) J^{a_k}(w_k)\ldots
   J^{a_1}(w_1)\rangle \notag  \allowdisplaybreaks \\
 &= (K_{b}^a)_i \oint_0\frac{\ud w_k}{2 \pi i w_k}\cdots\oint_0\frac{\ud w_1}{2 \pi i w_1}\oint_{z_i}\frac{\ud z}{2 \pi i (z-z_i)}
   \langle \phi_{s_1}(z_1)\ldots J^b(z)\phi_{s_i}(z_i) \ldots\phi_{s_N}(z_N) J^{a_k}(w_k)\ldots
   J^{a_1}(w_1)\rangle \notag  \allowdisplaybreaks \\
 &=-(K_{b}^a)_i \sum_{j(\neq i)} \oint_0\frac{\ud w_k}{2 \pi i w_k}\cdots\oint_0\frac{\ud w_1}{2 \pi i w_1}\oint_{z_j}\frac{\ud z}{2 \pi i (z-z_i)}
   \langle \phi_{s_1}(z_1)\ldots J^b(z)\phi_{s_i}(z_i)
   \ldots\phi_{s_N}(z_N) J^{a_k}(w_k)\ldots J^{a_1}(w_1)\rangle \notag  \allowdisplaybreaks \\
 &\phantom{=}-(K_{b}^a)_i \sum_{q=1}^k \oint_0\frac{\ud w_k}{2 \pi i w_k}\cdots\oint_0\frac{\ud w_1}{2 \pi i w_1}\oint_{w_q}\frac{\ud z}{2 \pi i (z-z_i)}
   \langle \phi_{s_1}(z_1)\ldots J^b(z)\phi_{s_i}(z_i)
   \ldots\phi_{s_N}(z_N) J^{a_k}(w_k)\ldots J^{a_1}(w_1)\rangle.
\end{align}
Inserting the OPE between a primary field and a current operator (cf. Eq. \eqref{eq:ope-j-with-vertex-op}) and the OPE
between two current operators \cite{Knizhnik1984},
\begin{align}
  \label{eq:OPE-of-JJ}
  J^a(z) J^b(w) &\sim \frac{\delta_{ab}}{2 (z-w)^2} + i \vep_{abc} \frac{J^c(w)}{z-w},
\end{align}
and writing $\PhiString = \Vi{1} \dots \Vi{N}$, we get
\begin{align}
0 &=(K_{b}^a)_i \sum_{j(\neq i)} \oint_0\frac{\ud w_k}{2 \pi i w_k}\cdots\oint_0\frac{\ud w_1}{2 \pi i w_1}\oint_{z_j}\frac{\ud z}{2 \pi i (z-z_i)}
   \frac{t_j^b}{z-z_j}\langle \PhiString J^{a_k}(w_k)\ldots J^{a_1}(w_1)\rangle  \notag \allowdisplaybreaks \\
 &\phantom{=}-(K_{b}^a)_i \sum_{q=1}^k \oint_0\frac{\ud w_k}{2 \pi i w_k}\cdots\oint_0\frac{\ud w_1}{2 \pi i w_1}\notag\\& \quad\quad\times \oint_{w_q}\frac{\ud z}{2 \pi i (z-z_i)}
   \frac{\delta_{ba_q}}{2(z-w_q)^2}\langle \PhiString J^{a_k}(w_k)\ldots J^{a_{q+1}}(w_{q+1}) J^{a_{q-1}}(w_{q-1}) \ldots J^{a_1}(w_1)\rangle \notag  \allowdisplaybreaks \\
 &\phantom{=}-(K_{b}^a)_i \sum_{q=1}^k \oint_0\frac{\ud w_k}{2 \pi i w_k}\cdots\oint_0\frac{\ud w_1}{2 \pi i w_1} \oint_{w_q}\frac{\ud z}{2 \pi i (z-z_i)}
   \frac{i\vep_{ba_qc}}{z-w_q} \langle \PhiString J^{a_k}(w_k)\ldots J^{c}(w_q) \ldots J^{a_1}(w_1)\rangle \notag  \allowdisplaybreaks \\
 &=-(K_{b}^a)_i \sum_{j(\neq i)} \frac{t_j^b}{z_i-z_j} \oint_0\frac{\ud w_k}{2 \pi i w_k}\cdots\oint_0\frac{\ud w_1}{2 \pi i w_1} \langle \PhiString
   J^{a_k}(w_k)\ldots J^{a_1}(w_1)\rangle \notag  \allowdisplaybreaks \\
 &\phantom{=}+(K_{a_q}^a)_i \sum_{q=1}^k \oint_0\frac{\ud w_k}{2 \pi i w_k}\cdots\oint_0\frac{\ud w_1}{2 \pi i w_1}\frac{1}{2(w_q-z_i)^2}
   \langle \PhiString J^{a_k}(w_k)\ldots J^{a_{q+1}}(w_{q+1}) J^{a_{q-1}}(w_{q-1}) \ldots J^{a_1}(w_1)\rangle \notag  \allowdisplaybreaks \\
 &\phantom{=}- (K_{b}^a)_i \sum_{q=1}^k \oint_0\frac{\ud w_k}{2 \pi i w_k}\cdots\oint_0
   \frac{\ud w_1}{2 \pi i w_1} \frac{i\vep_{ba_qc}}{w_q-z_i} \langle \PhiString J^{a_k}(w_k)\ldots J^{c}(w_q) \ldots J^{a_1}(w_1)\rangle \notag  \allowdisplaybreaks \\
 &=-(K_{b}^a)_i \sum_{j(\neq i)} \frac{t_j^b}{z_i-z_j} \psi_{a_k\ldots a_1} +\sum_{q=1}^k\frac{(K_{a_q}^a)_i}{2z_i^2}
   \psi_{a_k\ldots a_{q+1}a_{q-1}\ldots a_1} \notag  \allowdisplaybreaks \\
 &\phantom{=}-\sum_{q=1}^k (K_{b}^a)_i i\vep_{ba_qc} \oint_0\frac{\ud w_k}{2 \pi i w_k}\cdots\oint_0\frac{\ud w_q}{2 \pi i w_q}
   \frac{1}{w_q-z_i} \langle \PhiString J^{a_k}(w_k)\ldots J^{c}(w_q)(J_{-1}^{a_{q-1}} \ldots J_{-1}^{a_1})(0)\rangle \notag  \allowdisplaybreaks \\
 &=-(K_{b}^a)_i \sum_{j(\neq i)} \frac{t_j^b}{z_i-z_j} \psi_{a_k\ldots a_1} +\sum_{q=1}^k\frac{(K_{a_q}^a)_i}{2z_i^2} \psi_{a_k\ldots a_{q+1}a_{q-1}\ldots a_1}
   -\sum_{q=1}^k\sum_{n=-1}^{q-1} (K_{b}^a)_i i\vep_{ba_qc} \notag  \allowdisplaybreaks \\
 &\phantom{=}\times \oint_0\frac{\ud w_k}{2 \pi i w_k}\cdots\oint_0\frac{\ud w_q}{2 \pi i w_q} \frac{w_q^{-n-1}}{w_q-z_i}
   \langle \PhiString J^{a_k}(w_k)\ldots J^{a_{q+1}}(w_{q+1})(J_n^{c}J_{-1}^{a_{q-1}} \ldots J_{-1}^{a_1})(0)\rangle \notag  \allowdisplaybreaks \\
 &=-(K_{b}^a)_i \sum_{j(\neq i)} \frac{t_j^b}{z_i-z_j} \psi_{a_k\ldots a_1} +\sum_{q=1}^k\frac{(K_{a_q}^a)_i}{2z_i^2}
   \psi_{a_k\ldots a_{q+1}a_{q-1}\ldots a_1} \notag  \allowdisplaybreaks \\
 &\phantom{=}+(K_{b}^a)_i \sum_{q=1}^k\sum_{n=-1}^{q-1} \frac{i\vep_{ba_qc}}{z_i^{n+2}} \langle \PhiString (J_{-1}^{a_k}\ldots
   J_{-1}^{a_{q+1}}J_n^{c}J_{-1}^{a_{q-1}} \ldots J_{-1}^{a_1})(0)\rangle
\end{align}
Multiplying this equation by $2z_i$ gives
\begin{align*}
  (K_{b}^a)_i \sum_{j(\neq i)} (w_{ij}+1) t_j^b \psi_{a_k\ldots a_1} &= \sum_{q=1}^k\frac{(K_{a_q}^a)_i}{z_i} \psi_{a_k\ldots a_{q+1}a_{q-1}\ldots a_1}\notag\\
  &\quad+2  (K_{b}^a)_i \sum_{q=1}^k\sum_{n=-1}^{q-1} \frac{i\vep_{ba_qc}}{z_i^{n+1}} \langle \PhiString (J_{-1}^{a_k}\ldots J_{-1}^{a_{q+1}}J_n^{c}J_{-1}^{a_{q-1}} \ldots J_{-1}^{a_1})(0)\rangle
\end{align*}
\end{widetext}
The application of the spin operator $T^a$ to $\Jpsi$
can be written in terms of the Levi-Civita symbol,
\begin{align}
  \label{eq:T-Jpsi-in-epsilon}
  T^b \Jpsi \notag & = \Braket{\Vi{1} \dots \Vi{N} (J^b_0 J^{a_k}_{-1} \dots J^{a_1}_{-1})(0)} \notag \allowdisplaybreaks \\
  &\quad= i \sum_{q=1}^N \vep_{ba_qc} \psi_{a_k \dots a_{q+1} c a_{q-1}\dots q_1}.
\end{align}
Using this result, noting that $(K_{b}^a)_i t_i^b=0$
and that the operator $\Co{a}{i}$ can be written as $\Co{a}{i} = (K^{a}_{b})_i \sum_{j (\neq i)} w_{ij} t^b_j$,
we get
\begin{align*}
  &\mathcal{C}_i^a \psi_{a_k\ldots a_1} \\
  &= \sum_{q=1}^k\frac{(K_{a_q}^a)_i}{z_i} \psi_{a_k\ldots a_{q+1}a_{q-1}\ldots a_1} \\
  &\quad +(K_{b}^a)_i T^b\psi_{a_k\ldots a_1} \\
  &\quad + 2 (K_{b}^a)_i \sum_{q=2}^k\sum_{n=0}^{q-1} \frac{i\vep_{ba_qc}}{z_i^{n+1}} \\
  &\quad\quad \times \langle \PhiString (J_{-1}^{a_k}\ldots J_{-1}^{a_{q+1}}J_n^{c}J_{-1}^{a_{q-1}} \ldots J_{-1}^{a_1})(0)\rangle.
\end{align*}
\section{Lattice Momentum of Wave Functions}
\label{SEC:Lattice-momentum}
Following Ref.~\onlinecite{Nielsen2011},
we define the lattice momentum operator as $p=-i\ln(\mathcal{T})$, where
\begin{equation}
\mathcal{T}=\mathcal{P}_{N,N-1}\mathcal{P}_{N-1,N-2}\ldots\mathcal{P}_{2,1}
\end{equation}
is the translation operator and $\mathcal{P}_{ij}=2t_i^at_j^a+1/2$ is the operator
that permutes the states of spin $i$ and spin $j$. Note that the momentum is defined modulo $2\pi$.

The wave function $\psi_0$ as given
in Eq. \eqref{eq:vertex-op-correlator} is equivalent to
\begin{align}
  \label{eq:psi-tilde}
  &\tilde{\psi}_0(s_1, \dots, s_N) \\
  &\quad= \delta_{\mathbf{s}} \prod_{p=1}^N e^{i \pi (p-1) (s_{p} + 1) / 2} \prod_{n < m} (z_n - z_m)^{(s_n s_m + 1)/ 2}\notag,
\end{align}
because $\tilde{\psi}_0$ and $\psi_0$ differ only by a spin independent constant.

Applying $\mathcal{T}$ to $\tilde{\psi}_0$ gives
\begin{align}
  \mathcal{T}\tilde{\psi}_0&=\delta_\mathbf{s}\prod_{p=1}^Ne^{i\pi(p-1)(s_{p-1}+1)/2} \\
  &\quad \times \prod_{n=1}^{N-1}\prod_{m=n+1}^N(z_n-z_m)^{(s_{n-1}s_{m-1}+1)/2} \allowdisplaybreaks \notag \\
  &=(-1)^{N/2}\tilde{\psi}_0 \notag
\end{align}

Therefore the momentum of $\psi_0$ is
\begin{equation}
p_0=\left\{\begin{array}{cl}
\pi & \textrm{for } N/2 \textrm{ odd}\\
0 & \textrm{for } N/2 \textrm{ even}\\
\end{array}\right.
\end{equation}

With $\mathcal{T} \It{a}{-1} \mathcal{T}^{-1} = e^{-2 \pi i / N} \It{a}{-1}$, it follows
that $\Jpsi$ has the momentum $p_0 - 2 \pi k / N$.

\bibliography{refs}

\end{document}